\documentclass[apj]{emulateapj}

\usepackage[colorlinks=true,urlcolor=blue,citecolor=blue,linkcolor=blue]{hyperref}

\usepackage{graphicx}
\usepackage{epsfig}
\usepackage{longtable}
\usepackage{lscape}
\usepackage{natbib}
\usepackage{hyperref}
\usepackage{rotating,bm}
\renewcommand{\vec}[1]{\bm{#1}}

\shorttitle{Scattering polarization of Ly\,$\alpha$ in a 3D model of the chromosphere-corona transition region}
\shortauthors{\v{S}t\v{e}p\'an et al.}

\begin{document}

\title{Three-dimensional radiative transfer simulations of the scattering polarization of the hydrogen Ly\,$\alpha$ line in a MHD model of the chromosphere-corona transition region}

\author{
J. \v{S}t\v{e}p\'an\altaffilmark{1}, 
J. Trujillo Bueno\altaffilmark{2,3,4}, 
J. Leenaarts\altaffilmark{5},
and M. Carlsson\altaffilmark{6},
}
\altaffiltext{1}{Astronomical Institute ASCR, Fri\v{c}ova 298, 251\,65 Ond\v{r}ejov, Czech Republic}
\altaffiltext{2}{Instituto de Astrof\'isica de Canarias, E-38205 La Laguna, Tenerife, Spain}
\altaffiltext{3}{Universidad de La Laguna, Departamento de Astrof\'isica, 38206 La Laguna, Tenerife, Spain}
\altaffiltext{4}{Consejo Superior de Investigaciones Cient\'ificas, Spain}
\altaffiltext{5}{Institute for Solar Physics, Department of Astronomy, Stockholm University, AlbaNova University Centre, SE-106 91 Stockholm, Sweden}
\altaffiltext{6}{Institute of Theoretical Astrophysics, University of Oslo, P.O. Box 1029 Blindern, N-0315 Oslo, Norway}

\begin{abstract}
Probing the magnetism of the upper solar chromosphere requires  
measuring and modeling the scattering polarization produced by  
anisotropic radiation pumping in UV spectral lines. Here we apply  
PORTA (a novel radiative transfer code) to investigate the hydrogen  
Ly\,$\alpha$ line in a 3D model of the solar atmosphere resulting from  
a state of the art MHD simulation. At full spatial resolution the  
linear polarization signals are very significant all over the solar  
disk, with a large fraction of the field of view showing line-center  
amplitudes well above the 1\,\% level. Via the Hanle effect the  
line-center polarization signals are sensitive to the magnetic field  
of the model's transition region, even when its mean field strength is  
only 15 G. The breaking of the axial symmetry of the radiation field  
produces significant forward-scattering polarization in Ly\,$\alpha$,  
without the need of an inclined magnetic field. Interestingly, the  
Hanle effect tends to decrease such  forward-scattering polarization  
signals in most of the points of the field of view. When the spatial  
resolution is degraded, the line-center polarization of Ly\,$\alpha$  
drops below the 1\,\% level, reaching values similar to those  
previously found in 1D semi-empirical models (i.e., up to about  
0.5\,\%). The center to limb variation of the spatially-averaged  
polarization signals is qualitatively similar to that found in 1D  
models, with the largest line-center amplitudes at  
$\mu=\cos\theta\approx 0.4$ ($\theta$ being the heliocentric angle).  
These results are important, both for designing the needed space-based  
instrumentation and for a reliable interpretation of future  
observations of the Ly\,$\alpha$ polarization.
\end{abstract}



\section{Introduction\label{sec:intro}}

\begin{figure*}[t]
\centering
\begin{tabular}{cc}
\includegraphics[width=8.5cm]{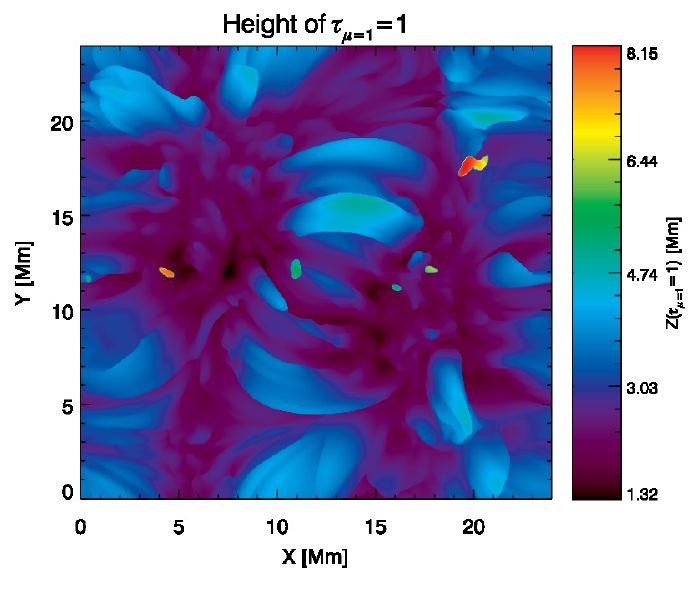} & \includegraphics[width=8.5cm]{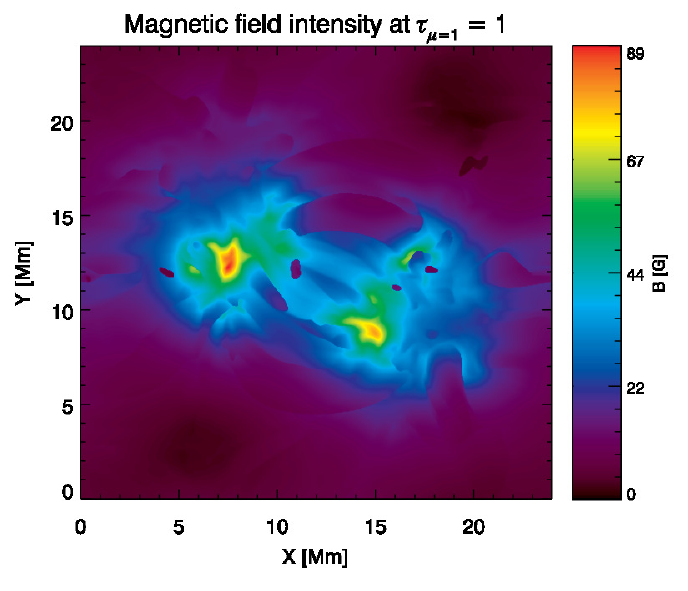}\\
\includegraphics[width=8.5cm]{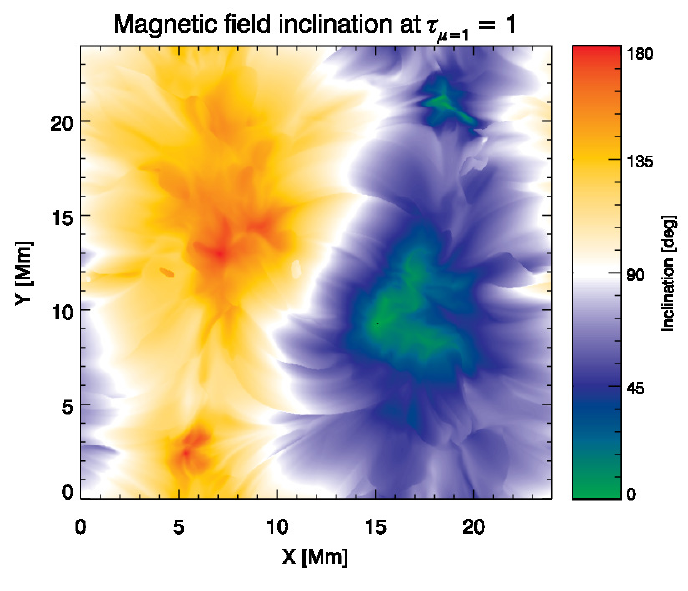} & \includegraphics[width=8.5cm]{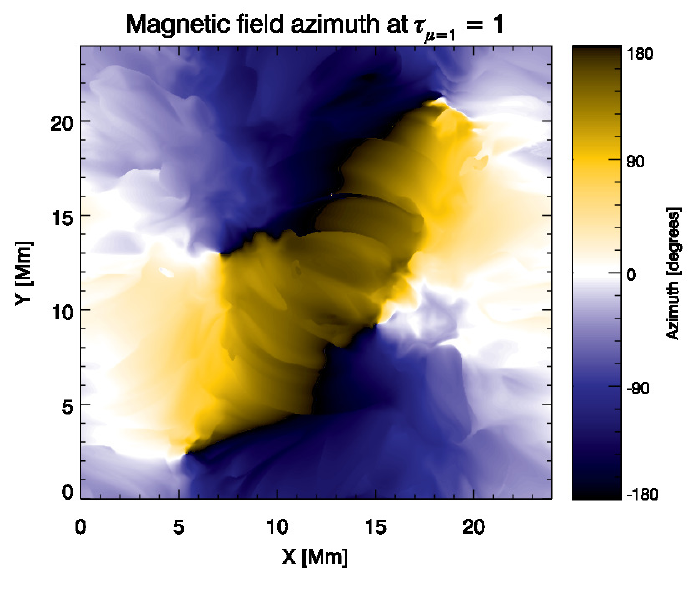}
\end{tabular}
\caption[]{
Top left panel: Horizontal variation of the height in the 3D model atmosphere where the Ly\,$\alpha$ line-center optical depth along the vertical direction is unity ($\tau_{\mu=1}=1$). Top right panel: Strength of the model's magnetic field at the atmospheric heights where $\tau_{\mu=1}=1$; the resulting average magnetic field strength is $\langle B\rangle\approx 15\,{\rm G}$. Bottom panels: Inclination (left) and azimuth (right) of the magnetic field at the heights where $\tau_{\mu=1}=1$. Note that between the two clusters of opposite polarity the magnetic field has a loop-like structure with a dominant azimuth and a strength varying between a few gauss and 60 gauss. 
}
\label{fig:mgf}
\end{figure*}

Spectrographs on board sounding rockets \citep[e.g.,][]{roussel-dupre-1982} and space telescopes \citep[e.g.,][]{curdt-2008} have successfully measured the intensity profile (Stokes $I(\lambda)$) of the hydrogen Ly\,$\alpha$ line radiation at various positions on the solar disk, showing its strong emission character. The center of the hydrogen Ly\,$\alpha$ emission profile originates at the base of the chromosphere-corona transition region (TR), where the kinetic temperature suddenly jumps from less than $10^4$ K to more than $10^6$ K and the plasma changes from partially to practically fully ionized. However, nobody has ever measured the linear polarization profiles ($Q(\lambda)$ and $U(\lambda)$) of the on-disk Ly\,$\alpha$ radiation. Recent radiative transfer investigations using one-dimensional (1D) and two-dimensional (2D) models of the solar atmosphere suggest that the absorption and scattering of anisotropic radiation in the upper solar chromosphere  produce measurable linear polarization in the hydrogen Ly\,$\alpha$ line radiation of the solar disk, in the line core and in the line wings \citep{jtb-stepan-casini11,belluzzi-trujillobueno-stepan,stepan12}. Moreover, via the Hanle effect the line-core polarization should react to the presence of magnetic fields in the solar TR, with sensitivity to field strengths between 10 and 100 G \citep{jtb-stepan-casini11,jtb-stepan-belluzzi12}.

The above-mentioned theoretical investigations on scattering polarization and the Hanle effect in the Ly\,$\alpha$ line of the solar-disk radiation have motivated the development of the Chromospheric Ly\,$\alpha$ Spectropolarimeter (CLASP), a sounding rocket experiment presently under development \citep{kobayashi12,kano12,kubo14}. The top priority of CLASP will be to measure, for the first time, the linear polarization pattern (i.e., $Q(\lambda)/I(\lambda)$ and $U(\lambda)/I(\lambda)$; hereafter, $Q/I$ and $U/I$) within a 1\,\AA\ wavelength interval around the Ly\,$\alpha$ line center. The second objective will be to use such observables to constrain the magnetic field of the upper solar chromosphere, by modeling the modification that the Hanle effect is expected to produce on the line-center linear polarization. The wing signals of the $Q/I$ profile can only be modeled taking into account partial frequency redistribution (PRD) and quantum interference between the sublevels pertaining to the two upper $J$-levels of the Ly\,$\alpha$ line \citep{belluzzi-trujillobueno-stepan}. Fortunately, as pointed out by \citet{belluzzi-trujillobueno-stepan}, the line-core signals of the $Q/I$ and $U/I$ profiles can be approximately modelled neglecting $J$-state interference and frequency correlations between the incoming and outgoing photons in the scattering events (i.e., within the limit of complete frequency redistribution, or CRD). This is important because the Hanle effect in Ly\,$\alpha$ operates only in the line core.

The solar atmosphere is highly inhomogeneous and dynamic. Plane-parallel, horizontally homogeneous atmospheric models can be used to estimate the significance of the scattering line polarization signals and their sensitivity to the Hanle effect \citep{jtb-stepan-casini11,jtb-stepan-belluzzi12}. However, for making progress in our understanding of the magnetic and thermal structure of the outer solar atmosphere it is important to confront spectropolarimetric observations with spectral synthesis calculations in increasingly realistic three-dimensional (3D) models resulting from state-of-the-art radiation magnetohydrodynamic (MHD) simulations, such as those 
carried out with the {\em Bifrost} code \citep{gudiksen11}. To make feasible such forward modeling research, \citet{porta} developed PORTA, a 3D multilevel radiative transfer code for modeling the intensity and polarization of spectral lines using massively parallel computers. 

\citet{stepan12} investigated the linear polarization of the hydrogen Ly\,$\alpha$ line produced by scattering processes in a two-dimensional (2D) model of the solar atmosphere. This 2D model was a slice from the {\em Bifrost} 3D model atmosphere used by \citet{leenaarts12} to investigate the intensity profiles of the H\,$\alpha$ line.   
Such 3D model of the extended solar atmosphere is representative of an enhanced network region and the aim of the present paper is to show in detail the results we have obtained by solving, with PORTA, the full 3D problem of the generation and transfer of scattering polarization in the Ly\,$\alpha$ line, taking into account the Hanle effect caused by the model's magnetic field. We do not show the polarization of the Zeeman effect because it produces rather negligible signals in Ly\,$\alpha$ \citep[e.g.,][]{jtb-14}. To study the full 3D problem of scattering polarization and the Hanle effect in chromospheric and transition region lines   
is important because 3D models of the solar atmosphere lack the translational symmetry of the previously investigated 1D and 2D cases, which provided only a glimpse on the complexity of the expected scattering polarization signals in the real 3D world.

This paper is organized as follows. After formulating the radiative transfer problem in Section~\ref{sec:formulation}, we show the radiation field tensors (Section~\ref{sec:radiation}) and the multipolar components of the atomic density matrix (Section~\ref{sec:atoms}) that result from the self-consistent solution in the chosen 3D model atmosphere. Section~\ref{sec:resolved} considers the spatially resolved case, showing the fractional polarization of the emergent Ly\,$\alpha$ radiation for two scattering geometries (disk center and close to the limb)  in the absence and in the presence of the model's magnetic field. In Section~\ref{sec:unresolved} we study the spatially averaged polarization signals, paying particular attention to their sensitivity to the spatial resolution element, to their center to limb variation, to their magnetic sensitivity via the Hanle effect, and to their detectability with a spectropolarimeter like CLASP. Section~\ref{sec:conclusions} summarizes our main conclusions.

\section{The radiative transfer problem}   \label{sec:formulation}

\begin{figure}[t]
\centering
\includegraphics[width=6cm]{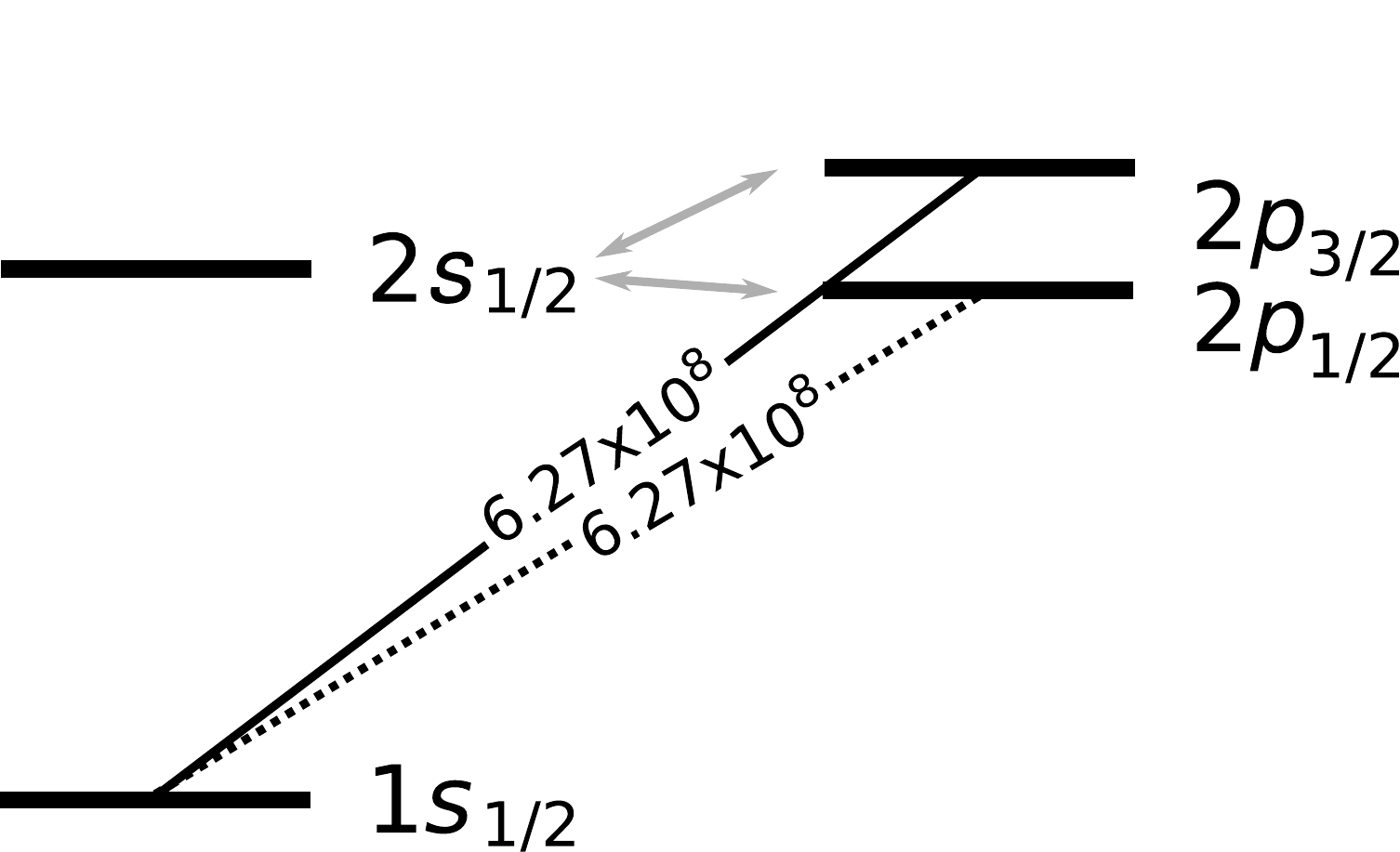}
\caption[]{Grotrian diagram of the hydrogen model atom. The components of Ly\,$\alpha$ are indicated along with their Einstein $A_{u\ell}$ coefficients (in $\mathrm{s}^{-1}$). The grey arrows connecting the fine-structure levels of $n=2$ indicate the weakly inelastic depolarizing transitions due to proton and electron collisions.}
\label{fig:grotr}
\end{figure}

\begin{figure*}[t]
\centering
\begin{tabular}{cc}
\includegraphics[width=8.5cm]{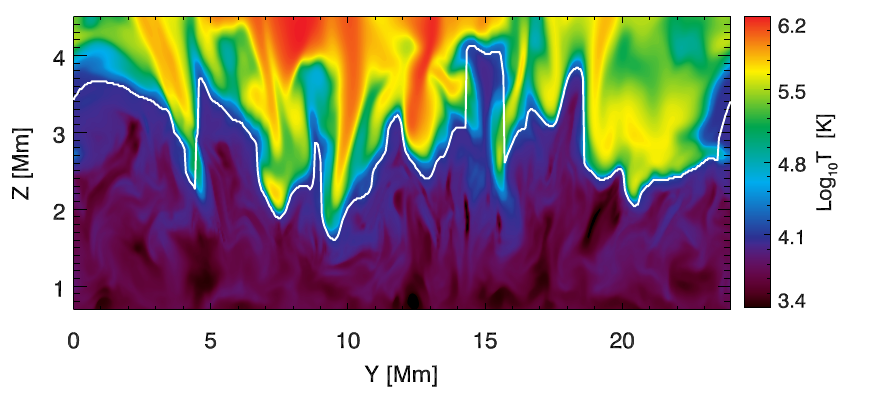}  & \includegraphics[width=8.5cm]{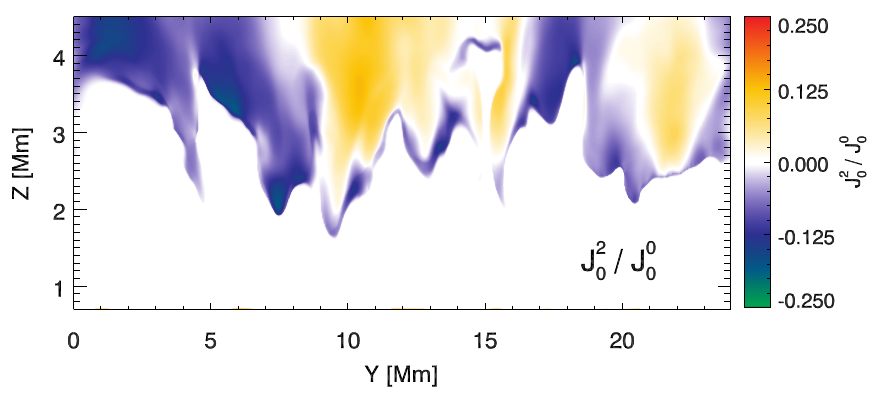} \\
\includegraphics[width=8.5cm]{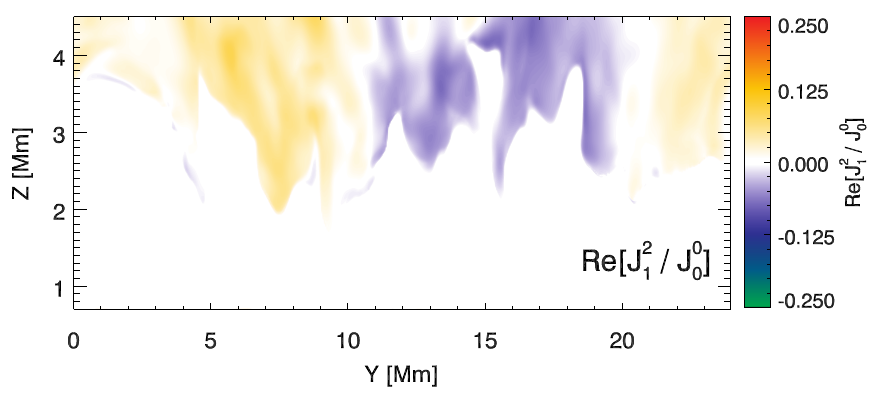} & \includegraphics[width=8.5cm]{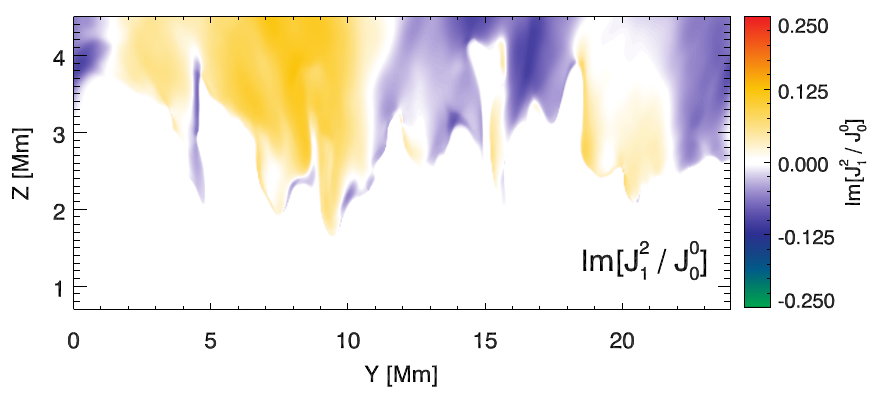} \\
\includegraphics[width=8.5cm]{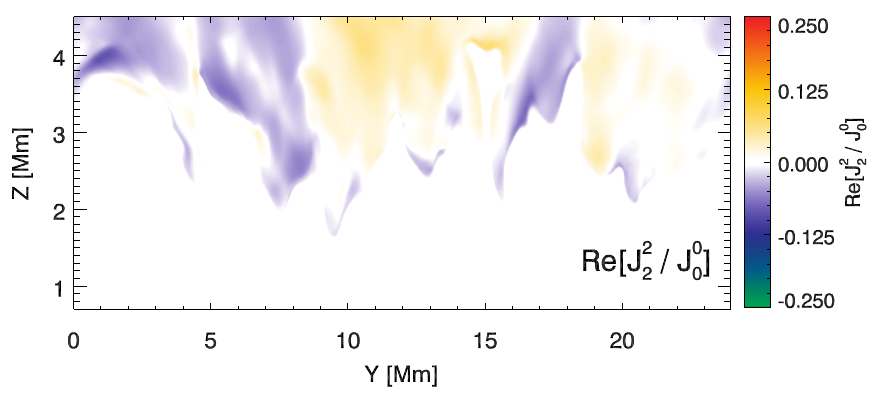} & \includegraphics[width=8.5cm]{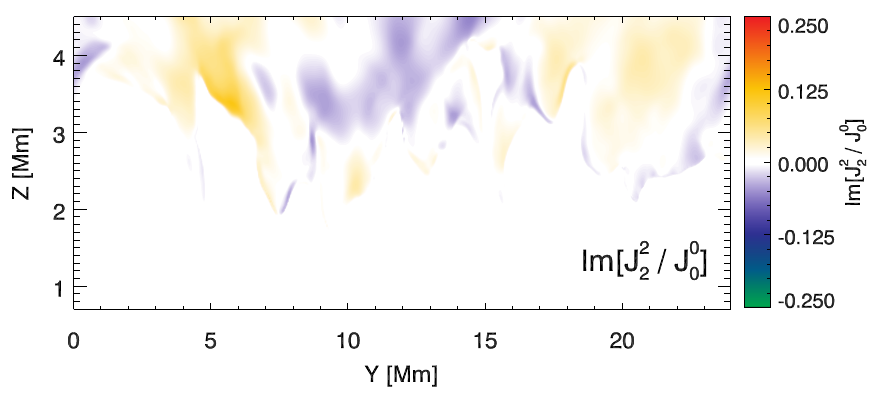} \\
\end{tabular}
\caption[]{
Physical quantities across the vertical 2D slice at $X=12$\,Mm in the 3D model. Upper left panel: Kinetic temperature. The white solid line indicates the height, associated to each horizontal Y-point, where the Ly\,$\alpha$ line-center optical depth is unity along a LOS with $\mu=1$. Other panels: Components of the radiation field tensors normalized to $J^0_0$.}
\label{fig:jkq}
\end{figure*}

\begin{figure}[t]
\centering
\begin{tabular}{c}
\includegraphics[width=8.5cm]{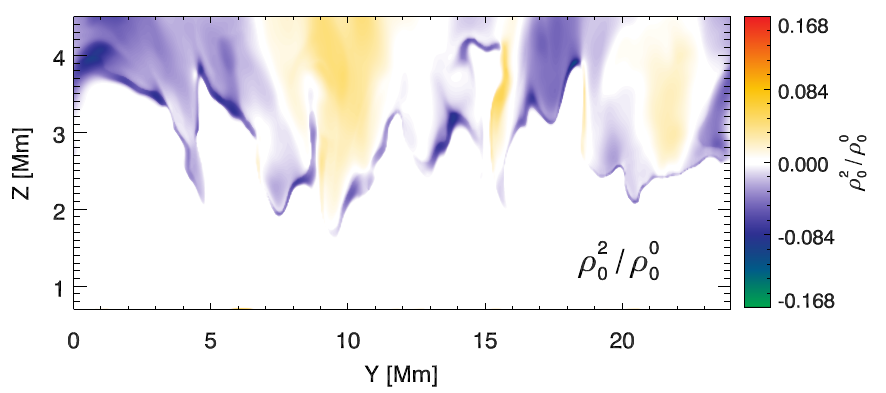} \\
\includegraphics[width=8.5cm]{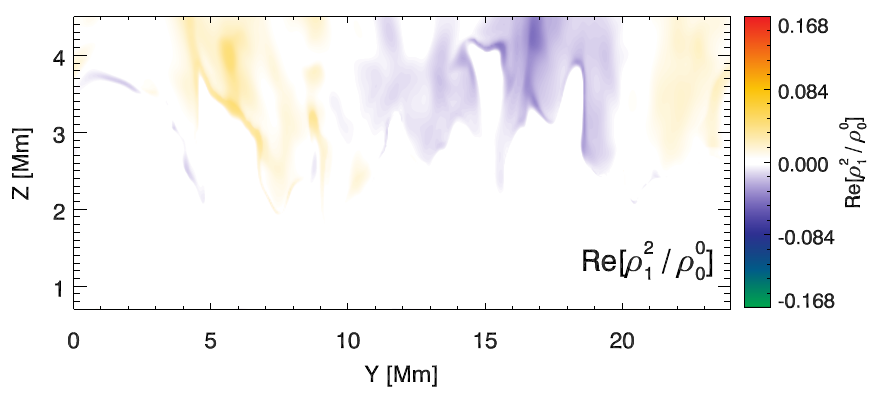} \\
\includegraphics[width=8.5cm]{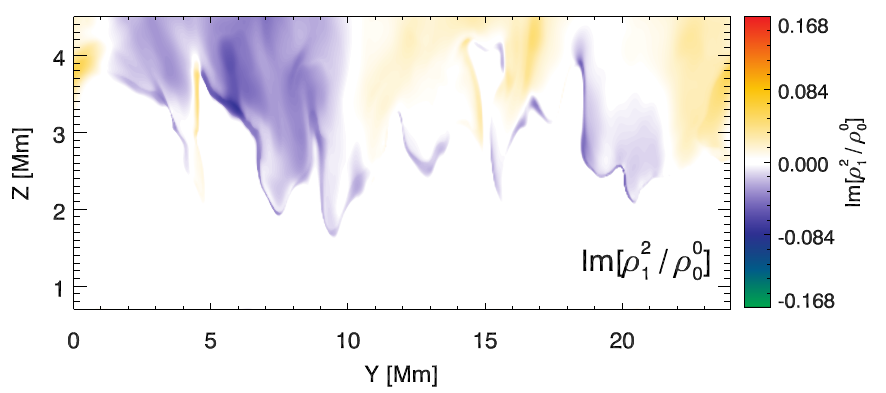} \\
\includegraphics[width=8.5cm]{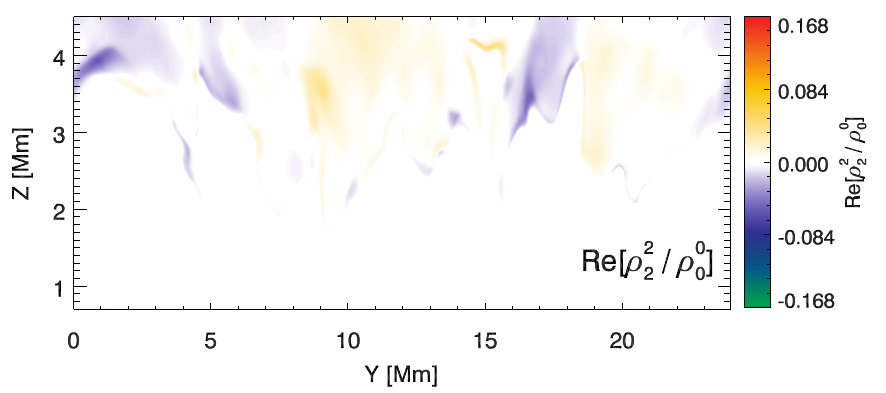} \\
\includegraphics[width=8.5cm]{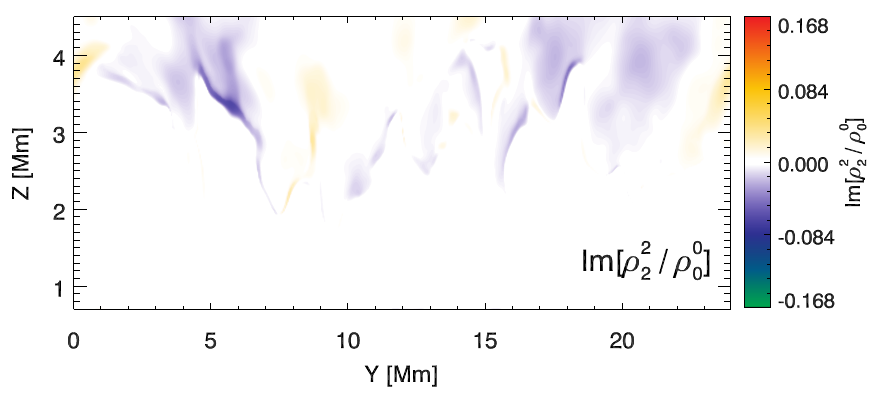} \\
\end{tabular}
\caption[]{
The fractional atomic alignment components, $\rho^2_Q/\rho^0_0$, of the $2p\,{}^2\! P_{\!3/2}$ upper level, 
in the same vertical slice as in Fig.~\ref{fig:jkq}.
}
\label{fig:rhokq}
\end{figure}

\begin{figure*}[t]
\centering
\begin{tabular}{cc}
\includegraphics[width=7cm]{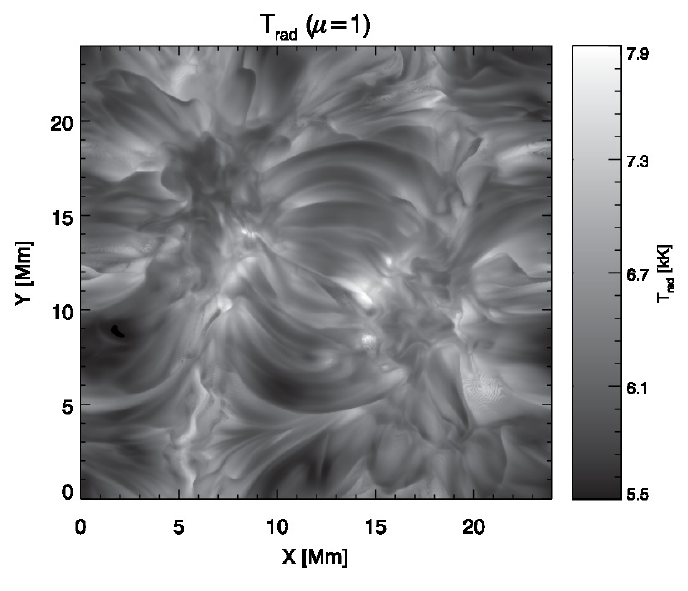} & \includegraphics[width=7cm]{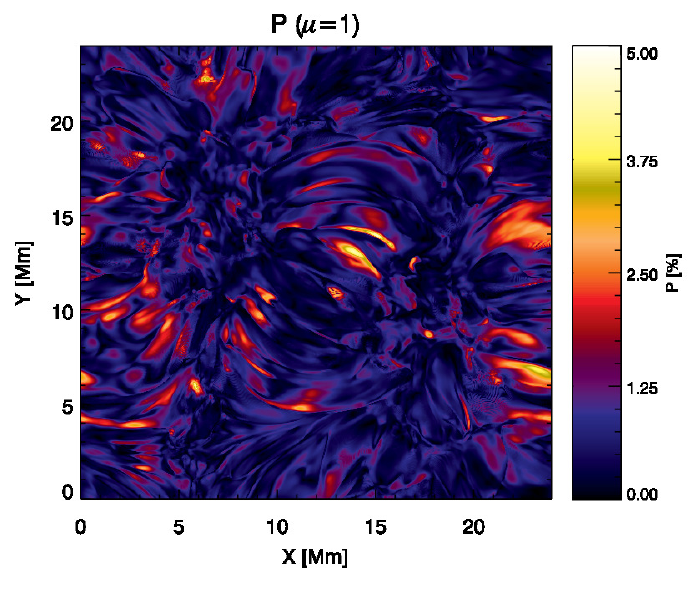} \\
\includegraphics[width=7cm]{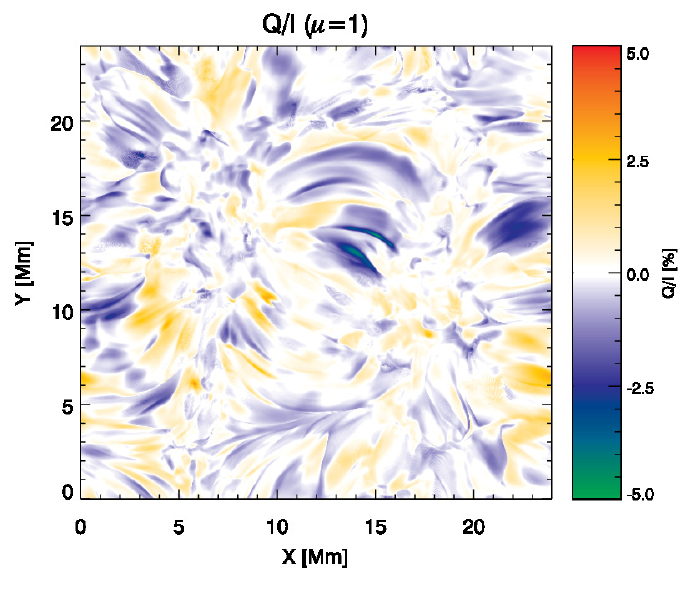} & \includegraphics[width=7cm]{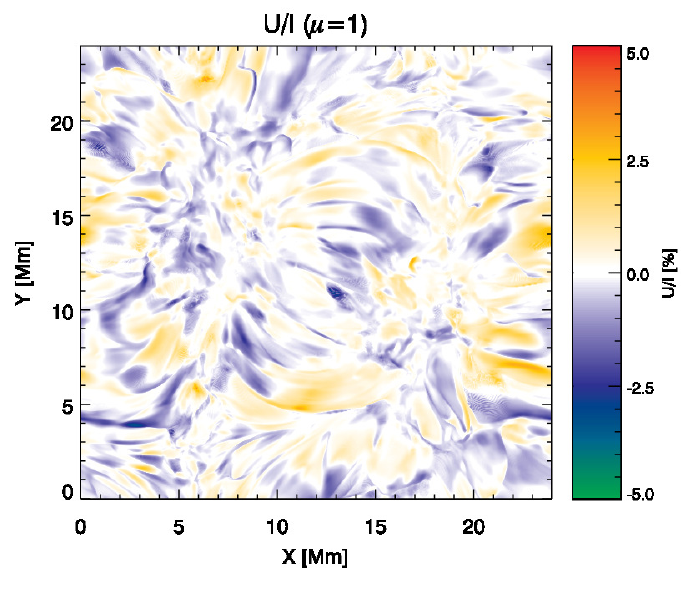} \\
\end{tabular}
\caption[]{
Synthetic disk center observation (LOS with $\mu=1$). Top left panel: the calculated emergent intensity at the Ly\,$\alpha$ line-center, quantified in terms of a radiation temperature. Top right panel: the total fractional linear polarization $P=\sqrt{Q^2+U^2}/I$ at the Ly\,$\alpha$ line center. Bottom panels: the corresponding $Q/I$ and $U/I$ line-center signals, with the positive reference direction for Stokes $Q$ along the $Y$-axis.
}
\label{fig:dco}
\end{figure*}

\begin{figure*}[t]  
\centering
\begin{tabular}{cc}
\includegraphics[width=7cm]{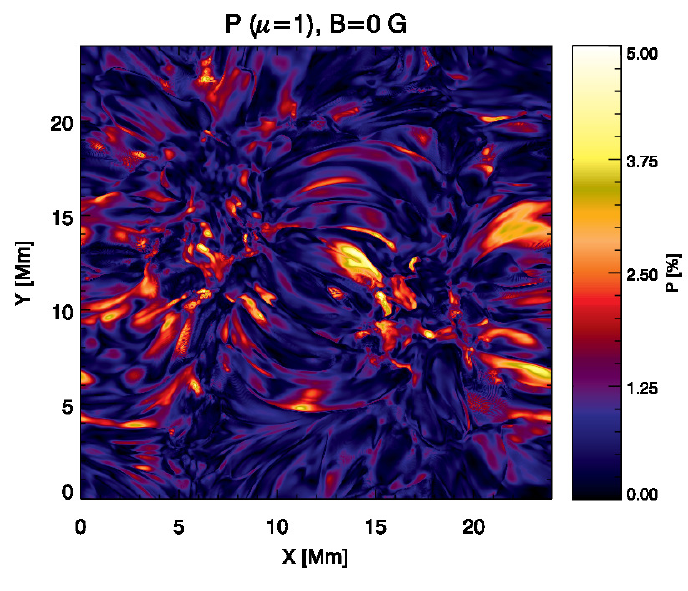} & \includegraphics[width=7cm]{fig_PL_mu100} \\ 
\includegraphics[width=7cm]{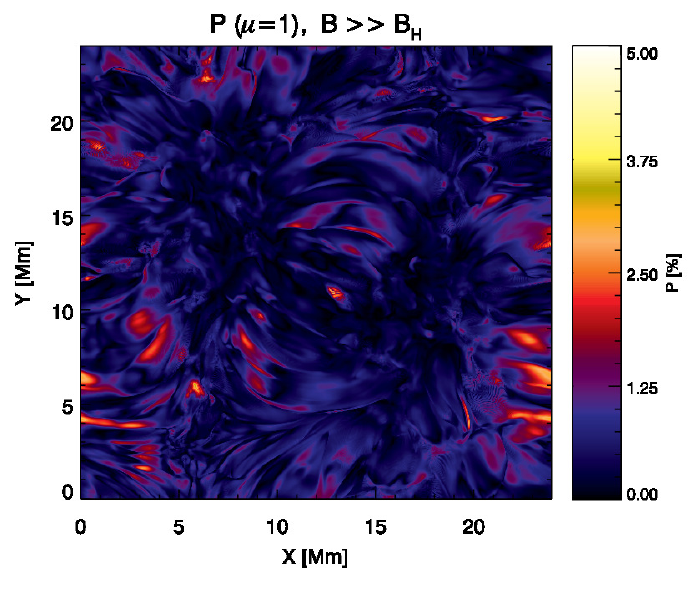} & \includegraphics[width=7cm]{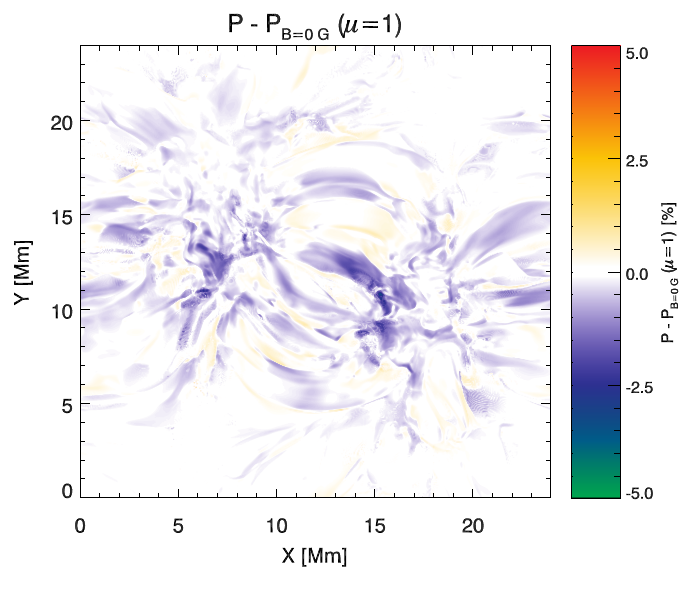} \\
\end{tabular}
\caption[]{
For the forward scattering case (LOS with $\mu=1$), the top panels show the total fractional linear polarization $P=\sqrt{Q^2+U^2}/I$ at the Ly\,$\alpha$ line center assuming the absence of magnetic fields (top left panel) and considering the Hanle effect produced by the model's magnetic field (top right panel). The bottom left panel shows the $P$ image that results when assuming that at each point within the 3D model the magnetic strength is in the Ly\,$\alpha$ line Hanle saturation regime (i.e., $B\gtrsim 250$\,G). The bottom right panel shows the difference between the true total fractional linear polarization (obtained by computing $I$, $Q$ and $U$ taking into account the Hanle effect of the model's magnetic field) and that corresponding to the zero-field reference case.
}
\label{fig:diff}
\end{figure*}

\begin{figure}[t]  
\centering
\includegraphics[width=8.5cm]{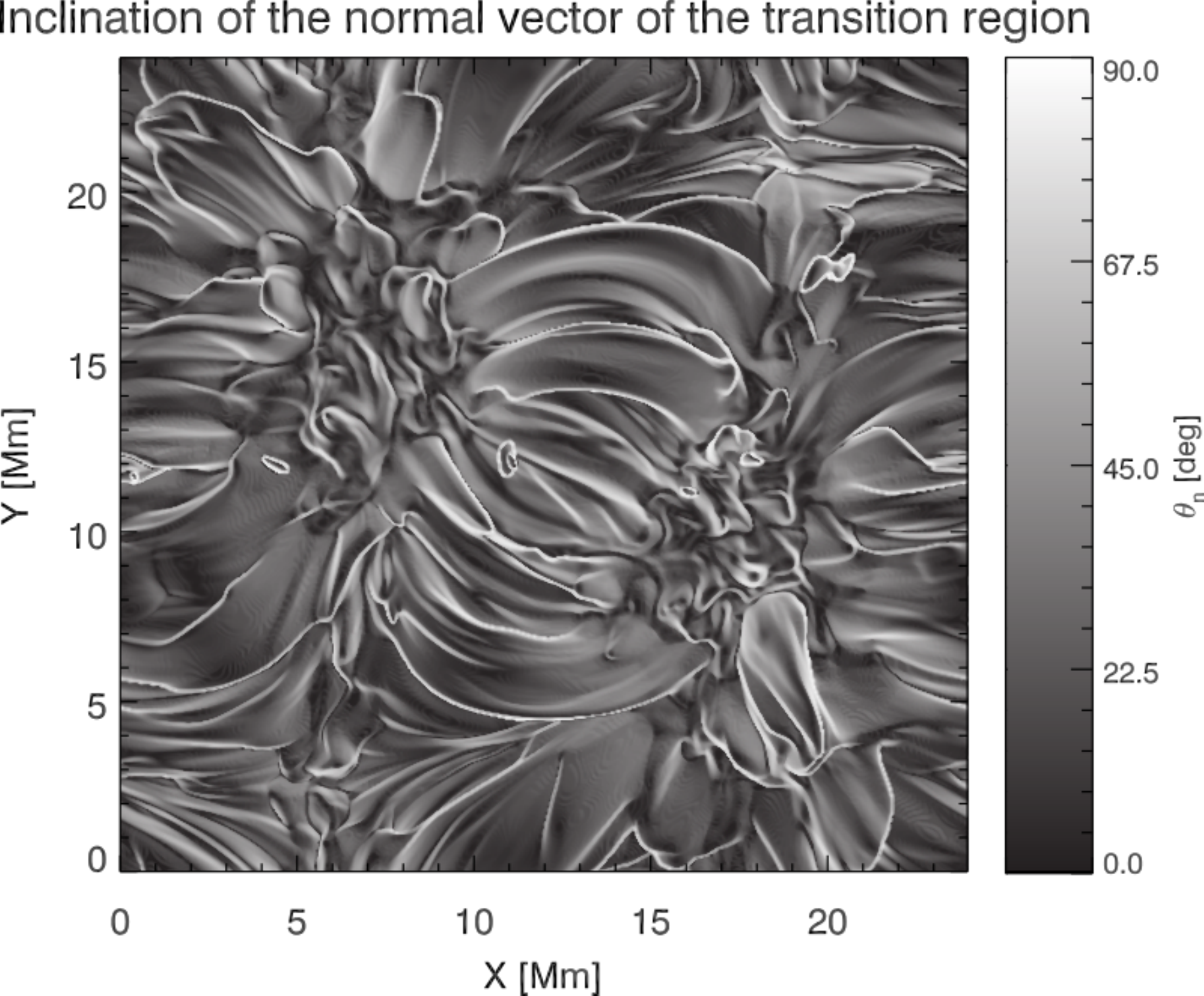}
\caption[]{
The inclination $\theta_{\rm n}$ of the vector that is normal to the corrugated surface that delineates the model's transition region, defined here as that where the Ly\,$\alpha$ line-center optical depth along the vertical direction is unity ($\tau_{\mu=1}=1$). The dark areas correspond to the places where the model's transition region is horizontal, as in a 1D model atmosphere, whereas the bright areas indicate the places where the TR is vertical. The average inclination is $\langle\theta_{\rm n}\rangle=26^\circ$. We note that a very similar figure is obtained if the normal vectors to the TR are defined as being parallel to the gradient of the kinetic temperature at the line-center formation heights.
}

\label{fig:trnorm}
\end{figure}

\begin{figure}[t]  
\centering
\begin{tabular}{cc}
\includegraphics[width=4.2cm]{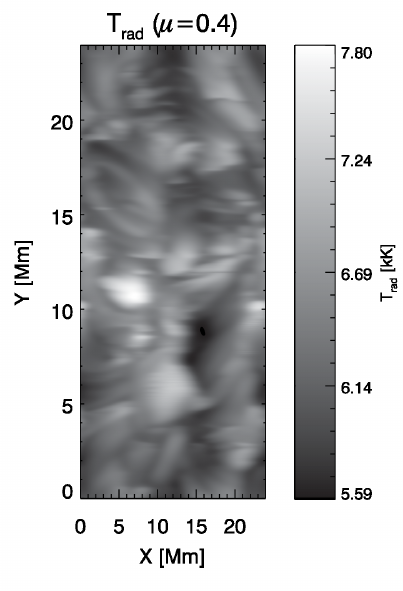} & \includegraphics[width=4.2cm]{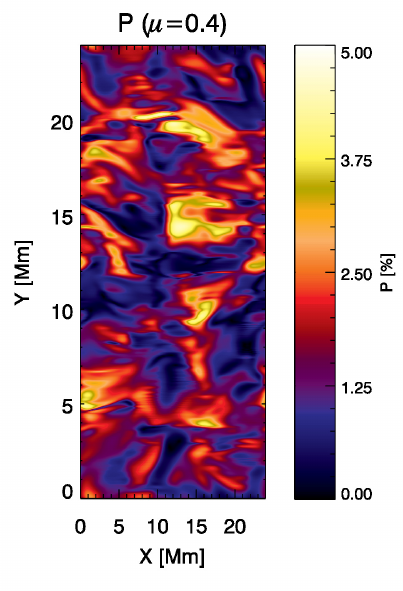} \\
\includegraphics[width=4.2cm]{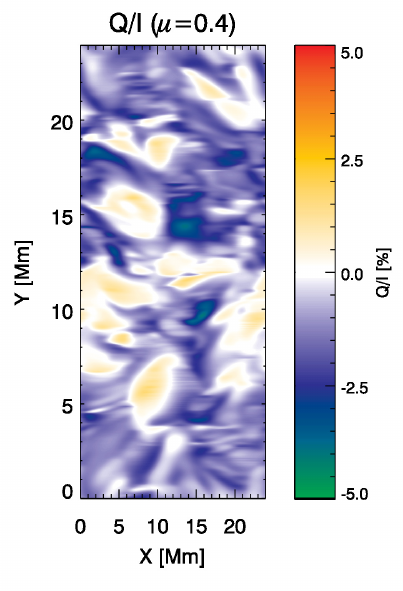} & \includegraphics[width=4.2cm]{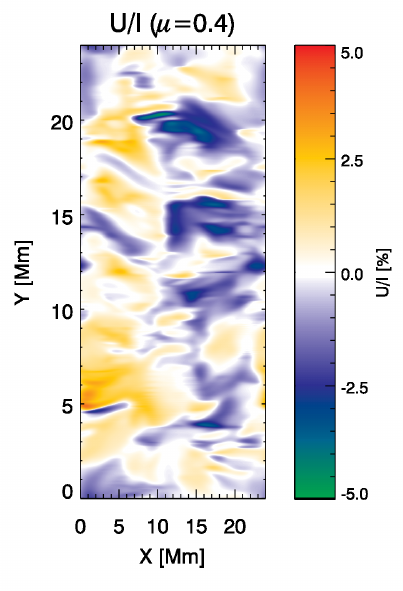} \\
\end{tabular}
\caption[]{
The same as in Fig.~\ref{fig:dco}, but for the case of a close to the limb observation ($\mu=0.4$). The visualization 
takes into account the foreshortening effect corresponding to the chosen inclined line of sight.
}
\label{fig:clo}
\end{figure}

\begin{figure}[t]
\centering
\begin{tabular}{cc}
\includegraphics[width=8cm]{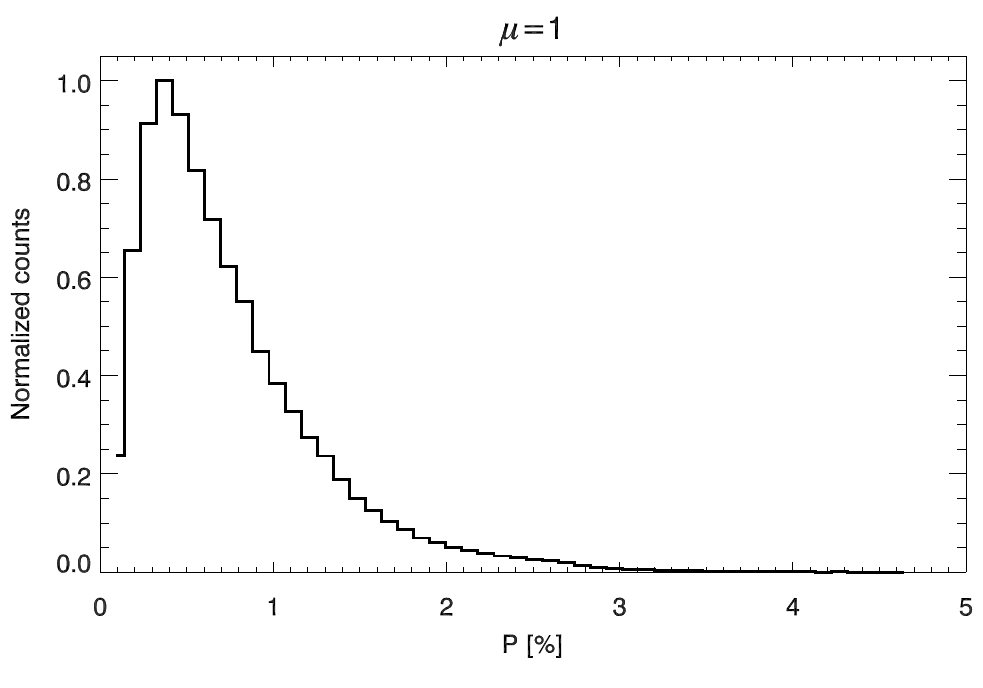} \\
\includegraphics[width=8cm]{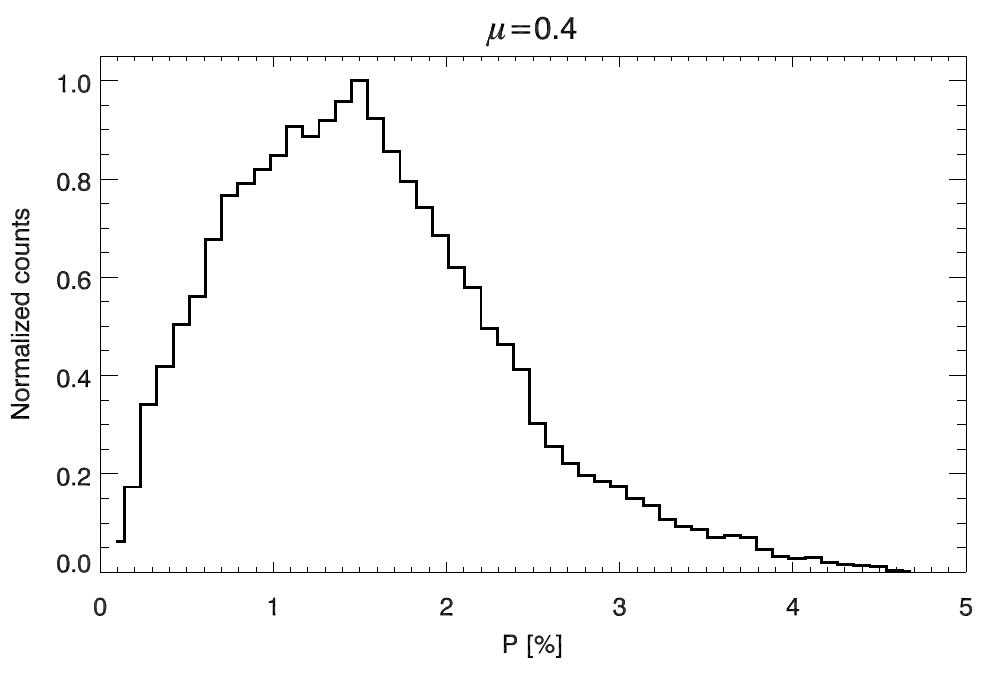} \\
\end{tabular}
\caption[]{
Histogram of the total fractional linear polarization $P=\sqrt{Q^2+U^2}/I$ at the Ly\,$\alpha$ line center.
Top panel: Disk center case (LOS with $\mu=1$). Bottom panel: close to the limb case (LOS with $\mu=0.4$).
}
\label{fig:histo}
\end{figure}

\begin{figure*}[t]
\centering
\begin{tabular}{cc}
\includegraphics[width=8cm]{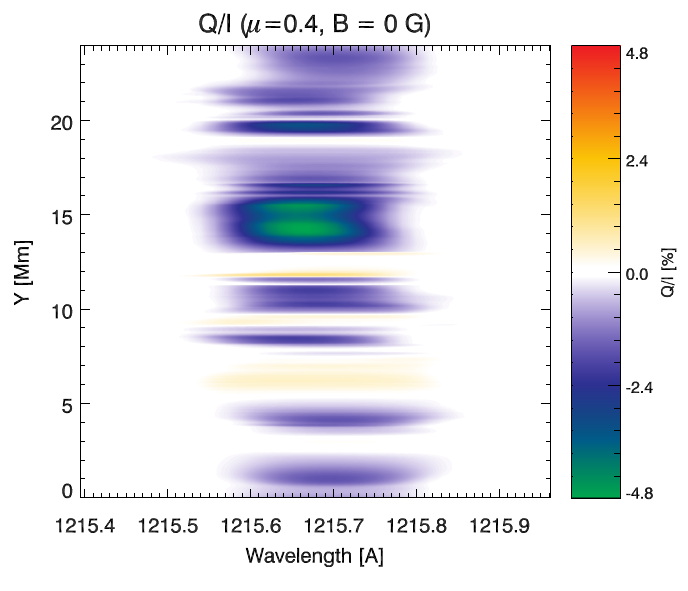} & \includegraphics[width=8cm]{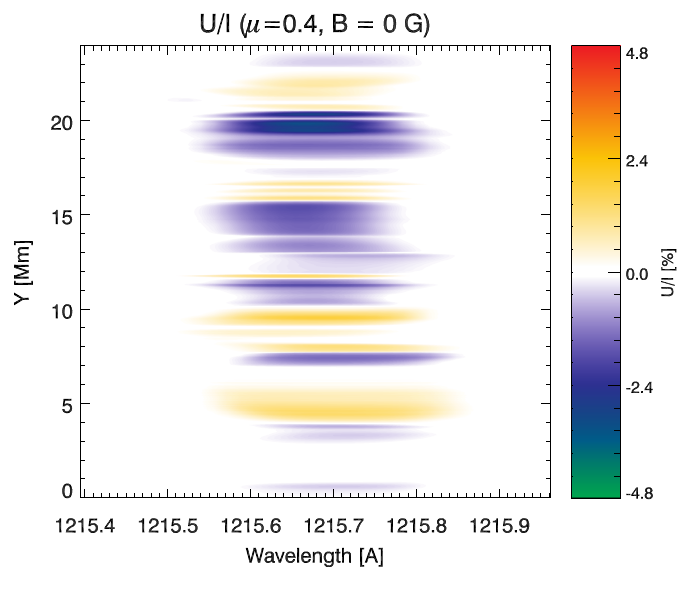} \\
\includegraphics[width=8cm]{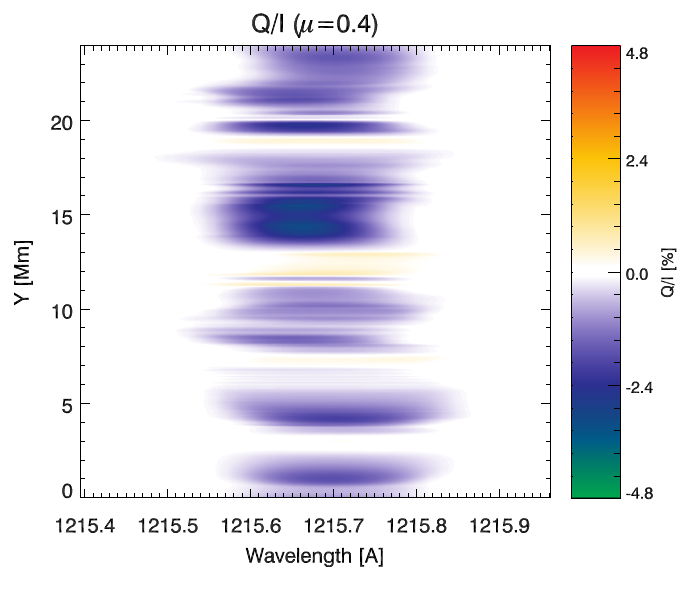} & \includegraphics[width=8cm]{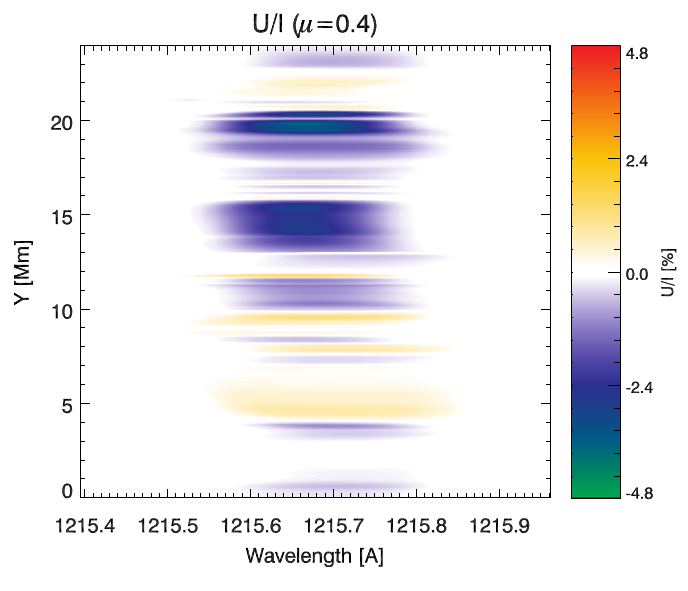} \\
\end{tabular}
\caption[]{
Simulated close to the limb observation ($\mu=0.4$). Emergent $Q/I$ and $U/I$ profiles calculated at each point along the $Y$-axis direction at $X$=12 Mm in Fig. 1. They have been obtained by solving the full 3D radiative transfer problem, neglecting (top panels) and taking into account (bottom panels) the Hanle effect produced by the model's magnetic field. The positive reference direction for Stokes $Q$ is parallel to the nearest limb (i.e., along the $Y$-axis).
}
\label{fig:clprofs}
\end{figure*}

\begin{figure}[t]
\centering
\includegraphics[width=8cm]{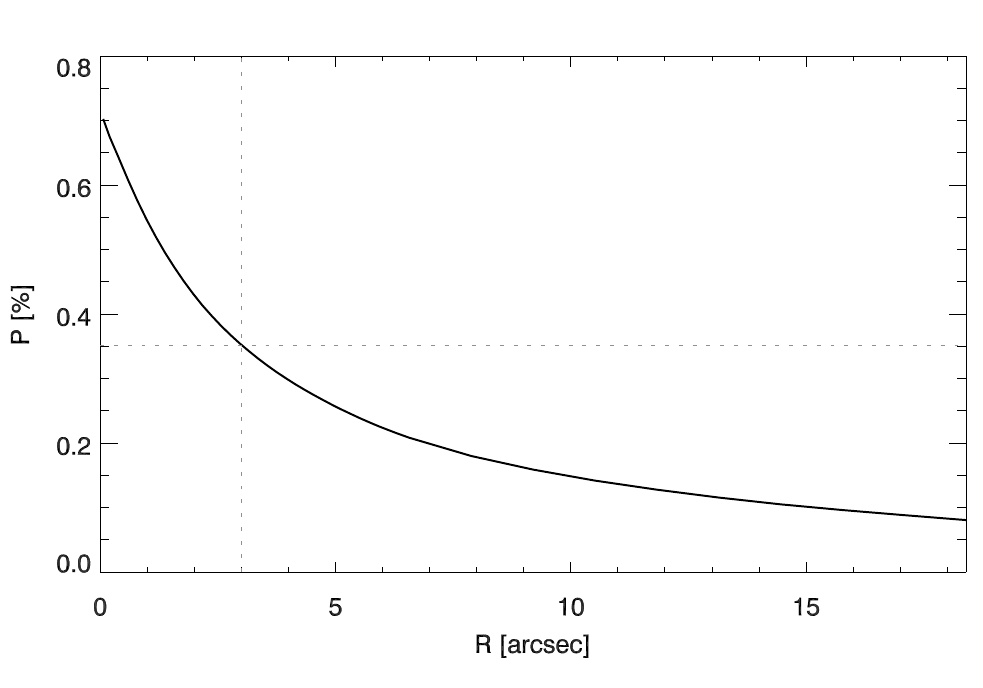}
\caption[]{
Average total fractional polarization signal $P$ for the forward-scattering case of a 
disk-center observation as a function of the spatial resolution $R$ (in arcseconds) of a square $R\times R$ element. The dotted lines indicate the resolution for which the average polarization signal is reduced by a factor $1/2$ with respect to the full-resolution case.
}
\label{fig:char}
\end{figure}

The 3D model atmosphere considered in this paper is a snapshot of a radiation MHD simulation performed with the {\em Bifrost} code \citep{gudiksen11} taking into account non-equilibrium hydrogen ionization \citep[see details in][]{carlsson-2014}. We used snapshot 385 of the simulation ``en024048-hion", which is the same simulation recently made publicly available through the Interface Region Imaging Spectrograph project \citep[IRIS;][]{depontieu-14} at the Hinode Science Data Center Europe (http://sdc.uio.no; see IRIS Technical Note 33). This 3D model atmosphere, which has horizontal periodic boundary conditions, 
encompasses the upper part of the convection zone, the photosphere, chromosphere, transition region and corona, and is identical to  that used by \citet{leenaarts12} to investigate the H$\alpha$ intensity profile. The model's magnetic field configuration is representative of an enhanced network region; it has magnetic field lines connecting two clusters of photospheric magnetic concentrations with two dominant opposite polarity regions 8 Mm apart, which reach chromospheric and coronal heights (see Fig.~\ref{fig:mgf}). 

The Ly\,$\alpha$ line results from two blended transitions between the ground level of hydrogen, $1s\,{}^2\! S_{\!1/2}$, and the $2p\,{}^2\!P_{\!1/2}$ and $2p\, {}^2\!P_{\!3/2}$ excited levels. The Ly\,$\alpha$ scattering polarization calculations of this paper were carried out fixing the model's neutral hydrogen number density at each spatial grid point (which were computed with the {\em Bifrost} code taking into account time-dependent hydrogen ionization) and solving the problem of the generation and transfer of polarized radiation assuming statistical equilibrium in a hydrogen model atom with only bound-bound transitions. The {\em Bifrost} ionization balance for hydrogen was computed using the method described in \citet{leenaarts07}, which assumes that the Lyman transitions are in detailed balance. 

The hydrogen model atom we have used in the scattering polarization calculations (see Fig.~\ref{fig:grotr}) is composed of the above-mentioned levels and the $2s\,{}^2\! S_{\!1/2}$ level. The collisional coupling of this level with the $P$-levels may contribute to a slight depolarization of the Ly\,$\alpha$ line in regions where the density of perturbers (protons and electrons) is significantly larger than $10^{11}\,{\rm cm}^{-3}$ \citep{stepanjtb11a}. Since in models of the upper solar chromosphere the proton and electron density is typically smaller than $10^{11}\,{\rm cm}^{-3}$, even a three-level model atom (i.e., without the $2s\,{}^2\! S_{\!1/2}$ level of Fig.~\ref{fig:grotr}) could be sufficiently realistic for computing the line-center polarization of Ly\,$\alpha$. The only level that produces scattering polarization in the core of the Ly\,$\alpha$ line is the $2p\,{}^2\! P_{\!3/2}$ upper level. It is interesting to note that the suitability of the above-mentioned three-level model atom implies that the atomic polarization of the $2p\,{}^2\! P_{\!3/2}$ level (population imbalances and quantum coherence between its magnetic sublevels) is dominated by the absorption of anisotropic radiation in the resonance line transition itself and by the action of the Hanle effect on such upper level, without any serious impact of collisional depolarization. The critical Hanle field of the level $2p\,{}^2\! P_{\!3/2}$ is $B_H{\approx}53$ G and the scattering polarization of the on-disk Ly\,$\alpha$ radiation is sensitive to magnetic strengths between 10 and 100 G, approximately \citep[]{jtb-stepan-casini11}

To compute the emergent Stokes profiles $I$, $Q$ and $U$ at each surface point of the atmospheric model we have to solve the following radiative transfer equations for any desired line of sight ($X$ being $I$, $Q$ or $U$)
\begin{equation}
{{d}\over{d{\tau}}}{X}=S_X - X\,,  
\label{eq:rte}
\end{equation}
where $d{\tau}={\eta_I}ds$ defines the monochromatic optical path length along the ray direction (with $s$ the corresponding geometrical distance), $\eta_I$ is the absorption coefficient, and $S_X=\epsilon_X/\eta_I$ is the source function of the Stokes parameter $X$, with $\epsilon_X$ the emission coefficient.\footnote{Note that in our radiative transfer problem ${\eta_I}$ is the only non-zero coefficient of the propagation matrix of the Stokes-vector transfer equation, because the Zeeman splitting is negligible compared to the width of the hydrogen Ly\,$\alpha$ line and the angular momentum of its lower level is $j=1/2$ and, therefore, cannot carry atomic alignment
\citep[i.e., we do not have ``zero-field" dichroism here; cf.][]{jtb-landi-07}.}
While the emission coefficient $\epsilon_I$ has contributions from the two blended transitions indicated in Fig.~\ref{fig:grotr}, $\epsilon_Q$ and $\epsilon_U$ depend only on the atomic polarization of the $2p\, {}^2\!P_{\!3/2}$ upper level (because it is the only Ly\,$\alpha$ level that can carry atomic alignment). 
We quantify the population and atomic polarization 
of this $j=3/2$ level by means of the multipolar components of its atomic density matrix (i.e., the $\rho^K_Q$ components, with $K=0,2$ and $Q=-K,...,K$, as quantified in the reference system whose quantization axis is along the local vertical) or, equivalently, by the following quantities (with the $\rho^K_Q$ components normalized to the fraction of neutral hydrogen atoms that are in the ground level): 
${S_Q^K}\,=\,{\frac{2h{\nu}^3}{c^2}}{\frac{2{j}_l+1}{\sqrt{2{j}_u+1}}}{\rho}_Q^K$.
In our radiative transfer calculations with PORTA we have used the exact $S_I$, $S_Q$ and $S_U$ expressions, which take into account that the two Ly\,$\alpha$ transitions are not perfectly overlapped. Such transitions are actually very close in wavelength (they  are only about 6 m\AA\ apart) and, for simplicity, we give here the $S_Q$ and $S_U$ expressions for the perfect overlapping case:


\begin{eqnarray}
S_Q&=&r\frac{\sqrt 2}{3}    \Big{\{} \frac{3}{2\sqrt{2}}(\mu^2-1) S^2_0 \label{eq:sq} \\
   &- & \sqrt{3}  \mu \sqrt{1-\mu^2} (\cos \chi
  {\rm Re}[{S}^2_1] - \sin \chi {\rm Im}[{S}^2_1])     \nonumber  \\
  &-&  \frac{\sqrt{3}}{2} (1+\mu^2) (\cos 2\chi  {\rm Re}[{S}^2_2]
  -\sin 2\chi \, {\rm Im}[{S}^2_2])   \Big{\}}\;, \nonumber
\end{eqnarray}
and 
\begin{eqnarray}
S_U&= \,r\,\sqrt{{2}\over{3}} \Big{\{} \sqrt{1-\mu^2} ( \sin \chi
  {\rm Re}[{S}^2_1]+\cos \chi {\rm Im}[{S}^2_1])     \label{eq:su}      \\ \nonumber
  &\quad+\mu (\sin 2\chi \, {\rm Re}[{S}^2_2] + \cos 2\chi \, {\rm Im}[{S}^2_2])\Big{\}}\;,
\end{eqnarray} 
where $\theta=\arccos \mu$ and $\chi$ are the inclination with respect to the solar local vertical and azimuth of the ray, respectively,
 and $r=\eta_I^{l}\phi_{\nu}/\eta_I=\eta_I^{l}\phi_{\nu}/(\eta_I^{l}\phi_{\nu}+\eta_I^{c})$ (with $\eta_I^{c}$ and $\eta_I^{l}$ the continuum and the line-integrated opacities, respectively, and $\phi_{\nu}$ the Voigt line absorption profile).
In such equations the reference
direction for Stokes $Q$ is in the plane formed by the 
ray's propagation direction and the vertical $Z$-axis. We point out that Eqs.~(\ref{eq:sq}) and (\ref{eq:su}),  
with their corresponding Eqs.~(\ref{eq:rte}), imply that the sources of 
Stokes $Q$ are the population imbalances ($S^2_0$) and the quantum coherence ($S^2_Q$, with $Q{\ne}0$), while the Stokes $U$ signal is produced only by quantum coherence ($S^2_Q$, with $Q{\ne}0$) between pairs of sublevels pertaining to the $2p\, {}^2\!P_{\!3/2}$ excited level.

To solve Eqs.~(\ref{eq:rte}) for any desired line of sight (LOS) we need to know first $\eta_I$ and the $S_X$ source function components at each spatial grid point. To obtain the self-consistent values of such  quantities we have to solve jointly the statistical equilibrium equations for the multipolar components ($\rho^K_Q(j)$) of the atomic density matrix corresponding to each atomic level $j$ and the radiative transfer equations for each of the radiative transitions in the chosen model atom. The general expressions of such equations can be found in Sections 7.2 and 7.13.c of \citet{LL04}, which we solved applying PORTA. We used a Gaussian quadrature for the ray inclinations with respect to the vertical $Z$-axis (with 5 inclinations per octant) and the trapezoidal rule for the ray azimuth (with 4 azimuths per octant). In our radiative transfer calculations we took into account the impact of the Doppler shifts, produced by the macroscopic velocity gradients of the 3D atmospheric model, on the anisotropic radiation pumping. Although this physical ingredient is not very important for modeling the scattering polarization of the hydrogen Ly\,$\alpha$ line (because the line is very broad), PORTA takes it fully into account because the dynamic state of the solar atmosphere may produce observable signatures on the scattering polarization of some chromospheric lines \citep{edgar2013}. Our calculations with PORTA were carried out in the MareNostrum III supercomputer of the Barcelona Supercomputing Center. The total computing time of the non-LTE calculations carried out for the investigation of this paper exceeded one million CPU hours.

To grasp better the physics of this complex radiative transfer problem it is useful to write down a simplified set of equations, instead of the general equations for the four-level model atom of Fig.~\ref{fig:grotr} we have actually solved with PORTA to obtain all the results shown in this paper.  To this end, we note that the weakly inelastic collisions with protons and electrons that couple the $2s\,{}^2\! S_{\!1/2}$ level with the $2p\, {}^2\!P_{{\!1/2}}$ and $2p\, {}^2\!P_{{\!3/2}}$ ones, which we have taken into account in our 3D radiative transfer calculations, do not really have a significant depolarizing effect on the Ly\,$\alpha$ polarization in the considered model atmosphere \citep{stepanjtb11a}. Taking also into account that elastic collisions with neutral hydrogen atoms can be safely neglected when modeling the line-core polarization of Ly\,$\alpha$ and that the only level contributing to the Ly$\alpha$ scattering polarization is the $2p\, {}^2\!P_{{\!3/2}}$ one, we have the following approximate expressions for the $S^K_Q$ tensors that quantify the excitation state of the $2p\, {}^2\!P_{{\!3/2}}$ upper level \citep{manso-jtb11}:


\begin{eqnarray}
S^0_0\,&=&\,(1-\epsilon){{J}}_0^0\,+\,{\epsilon}\,B_{\nu},  \label{eq:mtx}  \\ \nonumber \\
  \left( \begin{array}{l} \label{eq:mtx-20}
      {S^2_0} \\\\ 
      {{\rm Re}[{S}^2_1}] \\\\
      {{\rm Im}[{S}^2_1}] \\\\ 
      {{\rm Re}[{S}^2_2}] \\\\
      {{\rm Im}[{S}^2_2}]
    \end{array} \right) &=&
    {{(1-\epsilon)}\over{\sqrt{2}}} 
  \left( \begin{array}{r}
      {J}^2_0 \\\\
      {\rm Re}[{J}{}^{2}_{1}] \\\\
      -{\rm Im}[{J}{}^{2}_{1}] \\\\
      {\rm Re}[{J}{}^{2}_{2}] \\\\
      -{\rm Im}[{J}{}^{2}_{2}]
    \end{array} \right)\,
\\
    &\!\!\!\!\!\!\!\!\!\!\!\!\!\!\!\!\!\!\!\!\!\!\!\! +&\!\!\!\!\!\!\!\!\!\!\!\! {\Gamma_u}\, {{(1-\epsilon)}}\,
    \left( \begin{array}{cccccc}
        0 & M_{12} & M_{13} & 0 & 0 \\\\ 
        M_{21} & 0 & M_{23} & M_{24} & M_{25} \\\\
        M_{31} & M_{32} & 0 & M_{34} & M_{35} \\\\
        0 & M_{42} & M_{43} & 0 & M_{45} \\\\
        0 & M_{52} & M_{53} & M_{54} & 0 
      \end{array} \right) 
    \left( \begin{array}{l}
        {S^2_0} \\\\
        {{\rm Re}[{S}^2_1}] \\\\
        {{\rm Im}[{S}^2_1}] \\\\
        {{\rm Re}[{S}^2_2}] \\\\
        {{\rm Im}[{S}^2_2}]
     \end{array} \right)\,, \nonumber
\end{eqnarray}
where $B_{\nu}$ is the Planck function at the Ly\,$\alpha$ line-center frequency, 
$\epsilon$ is the photon collisional destruction probablility, ${\Gamma_u}=1.869\times10^{-2}B$ (with the magnetic strength $B$ in gauss), and the $M_{ij}$ coefficients of the magnetic kernel $\vec M$ depend on the inclination ($\theta_B$) of the local magnetic field vector with respect to the vertical $Z$-axis and on its azimuth ($\chi_B$). The frequency averaged (over the absorption profile $\phi_{\nu}$) radiation field tensors ${{J}_Q^K}$ that appear in such equations quantify the mean intensity (${{J}_0^0}$), the anisotropy (${{J}_0^2}$), and the breaking of the axial symmetry (the real and imaginary parts of ${{J}_1^2}$ and ${{J}_2^2}$) of the incident radiation field at each point within the medium; their explicit expressions can be found in \citet{manso-jtb11}, but note that the Stokes $I$ of the incident Ly\,$\alpha$ radiation has to be calculated taking into account that it results from two blended transitions (see Fig.~\ref{fig:grotr}). The equations for the ${S_Q^K}$ (atomic) quantities 
have a clear physical meaning. In particular, note that the magnetic operator $\vec M$ couples locally the $K=2$ components among them (the Hanle effect).

\section{The radiation field tensors} \label{sec:radiation}

The top left panel of Fig.~\ref{fig:jkq} shows the spatial variation of the kinetic temperature across a 2D slice through the above-mentioned 3D snapshot model atmosphere; it corresponds to the $Z$-$Y$ plane at $X=12$\,Mm in Fig.\ref{fig:mgf}. Note that the height of the TR changes drastically from one horizontal location to another, outlining a highly crumpled surface. As shown by the white solid curve, the corrugated surface that delineates the model's TR practically coincides with that defined by the Ly\,$\alpha$ line-center optical depth unity along the line of sight (LOS).

The other panels of Fig.~\ref{fig:jkq} show the radiation field tensors $J^K_Q$ of the Ly\,$\alpha$ radiation, normalized to ${J}^0_0$. Such quantities quantify the fractional anisotropy and the symmetry properties of the incident radiation at each point within the medium. We calculated them after obtaining with the 3D radiative transfer code PORTA the self-consistent solution in the 3D snapshot model illustrated in Fig.\,\ref{fig:mgf}, using the four-level model atom of Fig.~\ref{fig:grotr} and taking into account the (very weak) collisional coupling between the $2s\,{}^2\! S_{\!1/2}$ and the two $P$ levels.
Of particular interest is the spatial variation of the fractional anisotropy (top right panel); it is negligible in the photosphere and chromosphere of the model, but it suddenly becomes significant right at the location of the TR with negative values of about 10\%. The other radiation field tensors shown in the remaining panels quantify the breaking of the axial symmetry of the Ly\,$\alpha$ radiation at each point within the model atmosphere; note that at the atmospheric heights where the line center optical depth is unity such tensors may have sizable positive or negative values, depending on the horizontal point under consideration. 

\section{The atomic level polarization}  \label{sec:atoms}

The multipolar components of the atomic density matrix (i.e., the $\rho^K_Q$ quantities 
or, equivalently, the $S^K_Q$ ones) quantify the excitation state of the atomic level under consideration. As mentioned above, the scattering polarization of Ly\,$\alpha$ is produced by the atomic polarization of its $2p\, {}^2\!P_{\!3/2}$ upper level; therefore, in this section we provide information on the spatial variation of the level's fractional atomic polarization throughout the same $Z$-$Y$ plane considered in Fig.~\ref{fig:jkq}. Eqs.~(\ref{eq:mtx-20}) indicate that for $\Gamma_u=0$ (i.e., in the unmagnetized case) each multipolar component $\rho^2_Q$ of the $2p\, {}^2\!P_{\!3/2}$ level is directly proportional to its corresponding radiation field tensor component. We can say that there is a transfer of `order' from the radiation field to the atomic system, and Eqs.~(\ref{eq:rte})--(\ref{eq:su}) show that the atoms of a polarized (`ordered') atomic system emit linearly polarized radiation. Eqs.~(\ref{eq:mtx-20}) indicate that if all $J^2_Q$ were zero then all $S^2_Q$ would vanish, even in the presence of a magnetic field. Through the Hanle effect the magnetic field modifies the polarization of the atomic level according to the $\Gamma_u$ term of Eqs.~(\ref{eq:mtx-20}), and this has an impact on the emergent linear polarization $Q/I$ and $U/I$ signals. 

Fig.~\ref{fig:rhokq} shows the multipolar components of the density matrix corresponding to the $2p\, {}^2\!P_{\!3/2}$ level, normalized to the $\rho^0_0$ value of the same level 
(which is proportional to its overall population). 
Note that the $\rho^2_0/\rho^0_0$ panel (which quantifies the population imbalances among the sublevels of the $2p\, {}^2\!P_{\!3/2}$ level) shows mainly negative values at the atmospheric heights where the Ly\,$\alpha$ line center optical depth is unity along the LOS, while each of the quantum coherence shown in the other panels shows both positive and negative values. 

\section{The spatially resolved $Q/I$ and $U/I$ signals} \label{sec:resolved}

This section considers the case of spatially resolved observations. It shows maps of the line-center scattering polarization, which we have obtained by calculating the emergent Stokes profiles of the Ly\,$\alpha$ radiation after obtaining the self-consistent solution of the above-mentioned 3D radiative transfer problem. Of particular interest is to investigate the spatial variability of the emergent $Q/I$ and $U/I$ line-center signals and their sensitivity to the model's magnetic field. To this end, we solved the 3D radiative transfer problem of resonance line polarization for the following two cases: (1) taking into account the Hanle effect produced by the magnetic field of the 3D snapshot model atmosphere and (2) assuming $B=0$\,G at each spatial grid point of the 3D model. We show the results for two line of sights (LOS) defined by $\mu=\cos\theta$ (with $\theta$ the heliocentric angle): $\mu=1$ (corresponding to a disk center observation) and $\mu=0.4$ (corresponding to another on-disk observation, but at about 80~seconds of arc from the solar limb). In all our calculations of the emergent Stokes profiles the azimuth of the corresponding LOS was $\chi= 0^{\circ}$ -- that is, the LOS was contained in the $X$-$Z$ plane. In all the following figures we have chosen the $Y$-axis as the positive reference direction for Stokes $Q$, and for any LOS with $\mu<1$ we chose this $Y$-direction as the parallel to the nearest limb.

Fig.~\ref{fig:dco} shows, for the forward scattering case of a disk center observation (LOS with $\mu=1$), the intensity (top left panel), the total fractional linear polarization ($P=\sqrt{Q^2+U^2}/I$, top right panel), and the $Q/I$ (bottom left panel) and $U/I$ (bottom right panel) signals at the center of the emergent Ly\,$\alpha$ intensity profile. The calculated Ly\,$\alpha$ line-center intensity variations across the surface of the 3D atmospheric model show fibril-like structures, reminiscent of those observed at 1~arcsecond resolution by the VAULT sounding rocket \citep[e.g.,][]{vourlidas-2010}. Although we do not show illustrations here, we point out that the effects of 3D radiative transfer do have an impact on the emergent line-center intensities, as expected from the smoothing effect produced by horizontal radiative transfer under non-LTE conditions \citep[e.g.,][]{kneer81}; the so-called 1.5D approximation gives a larger line-center intensity contrast across the field of view\footnote{With this approximation, every column of the model atmosphere is approximated by a 1D plane-parallel model, so that horizontal radiative transfer effects are neglected.}. The same applies to other spectral lines such as hydrogen H$\alpha$ for which this effect is even more significant due to the lower opacity of the line \citep{leenaarts07}.

It is interesting to note that at high spatial resolution the linear polarization signals of the emergent Ly\,$\alpha$ line-center radiation are very significant (see Fig.~\ref{fig:dco}). At first sight this result might appear surprising, given that in Fig.~\ref{fig:dco} we are considering the forward scattering case of a disk-center observation. However, in this scattering geometry zero linear polarization should be expected only for the idealized case of an unmagnetized and horizontally homogeneous stellar atmosphere model. As soon as there is a magnetic field inclined with respect to the symmetry axis of the incident radiation field (hereafter, inclined field) and/or horizontal atmospheric inhomogeneities (e.g., in the gas kinetic temperature and/or density) the ensuing symmetry breaking effects can produce significant line scattering polarization signals \citep[see][and more references therein]{manso-jtb11,stepan12}. The linear polarization signals of Fig.~\ref{fig:dco} are due to the breaking of the axial symmetry of the Ly\,$\alpha$ radiation at each spatial point within the 3D model atmosphere, as quantified by the real and imaginary parts of the $J^2_1$ and $J^2_2$ radiation field tensors, as well to the magnetic field of the atmospheric model (see Eqs.~\ref{eq:mtx}--\ref{eq:mtx-20}). The $Q/I$ and $U/I$ line-center signals shown in the bottom panels of Fig.~\ref{fig:dco} have similar values that fluctuate in sign across the field of view, as expected from the particularization of Eqs.~(\ref{eq:sq}) and (\ref{eq:su}) to the case of a LOS with $\mu=1$. In particular, the image corresponding to the total linear polarization at the line center (see the top right panel of Fig.~\ref{fig:dco}) shows an impressive detail of fine-scale structuring.

As explained above, the forward-scattering polarization signals are caused by both the model's horizontal atmospheric inhomogeneities in temperature, density, and macroscopic velocity, and by the magnetic field. As shown in the top left panel of Fig.~\ref{fig:diff}, there are very significant forward scattering polarization signals in the absence of magnetic fields. Such zero-field linear polarization signals are solely due to the symmetry breaking caused by the horizontal atmospheric inhomogeneties; therefore, their very existence indicates  that the so-called 1.5D approximation is unsuitable for modeling high spatio-temporal resolution observations of 
the scattering polarization in Ly\,$\alpha$. Fig.~\ref{fig:diff} shows that there is a significant difference between the true total fractional linear polarization (calculated taking into account the Hanle effect produced by the magnetic field of the 3D model atmosphere) and that corresponding to the zero-field reference case. As expected, the action of the Hanle effect in forward scattering geometry is found mainly along the diagonal of the image seen in Fig.~\ref{fig:diff}, where the magnetization of the model atmosphere is the largest. However, as seen in the figure, the model's magnetic field tends to depolarize the emergent radiation, and this occurs mainly in the strongly polarized regions of the unmagnetized case. This contrasts with the case of a plane-parallel model atmosphere, where in the absence of magnetic fields the symmetry axis of the incident radiation coincides with the vertical direction and the Hanle effect of an inclined magnetic field creates linear polarization in forward scattering geometry but depolarization in close to the limb observations. In our 3D model atmosphere the symmetry axis of the radiation field that illuminates each of the points of the TR does not in general coincide with the local vertical. In Fig.~\ref{fig:trnorm}, we attempt to visualize the geometrical complexity of the present 3D problem by showing, at each point of the field of view, the inclination of the vector that is normal to the TR surface. In the unmagnetized case the strongly polarized regions are typically those for which the inclination of the local normal vector to the TR surface is larger, and depolarization by the Hanle effect is statistically more likely. In regions where the local normal vector to the TR surface is predominantly vertical the forward scattering signals of the unmagnetized case are typically lower and the Hanle effect produced by the model's magnetic field tends to be more similar to that found in the case of a plane-parallel model atmosphere. These two situations can be understood by analogy with the scattering polarization signals calculated in a 1D plane-parallel model near the limb and at the disk center, respectively \citep[cf.,][]{jtb-stepan-casini11}.

We now consider a LOS with $\mu=0.4$, which corresponds to the case of an observation relatively close to the solar limb. Fig.~\ref{fig:clo} shows maps of the line-center intensity (top left panel), of the total fractional linear polarization (top right panel), as well as of the $Q/I$ (bottom left panel) and $U/I$ (bottom right panel) line-center signals. The particularization of Eqs.~(\ref{eq:sq}) and (\ref{eq:su}) to the LOS under consideration (or to any LOS with $\mu{<}1$) indicates that $\epsilon_U$ (the emissivity in Stokes $U$ at each point along the LOS) depends only on the quantum coherence ($\rho^2_1$ and $\rho^2_2$) of the $2p\, {}^2\!P_{\!3/2}$ excited level, while $\epsilon_Q$ (the emissivity in Stokes $Q$ at each point along the LOS) depends also on the population imbalances between the sublevels of the excited level. As indicated by Eq.~(\ref{eq:mtx-20}), the population imbalances, quantified by $S^2_0$, are determined mainly by the radiation field tensor $J^2_0$, which at the crumpled surface that delineates the model's TR tends to be negative \citep[as in the 1D atmospheric models considered by][]{jtb-stepan-casini11}. We therefore expect that the $Q/I$ signals of the emergent Ly\,$\alpha$ radiation across the surface of the 3D model have mainly negative sign. For the $U/I$ signals we expect the occurrence of positive and negative values to be equally likely, following the fact (see Eq.~\ref{eq:mtx-20}) that the radiation field tensors $J^2_1$ and $J^2_2$ (which quantify the symmetry breaking of the incident radiation field) play a key role on the quantum coherence quantified by $S^2_1$ and $S^2_2$.

The upper panel of Fig.~\ref{fig:histo} shows the histogram of the line center polarization amplitudes of the  
emergent Ly$\alpha$ radiation calculated in forward scattering geometry (LOS with $\mu=1$) assuming $B=0$\,G. Most of the linear polarization signals have values between 0 and 1\,\%, but there is a clear tail of stronger polarization signals extending up to about 3\,\%. In this forward scattering geometry the mean fractional linear polarization amplitude at the maximum spatial resolution of the 3D model is $\overline{P}_{\mu=1}\approx 0.7\,\%$

The bottom panel of Fig.~\ref{fig:histo} shows an analogous histogram, but for the case of a LOS with $\mu=0.4$.  As in the $\mu=1$ case (see the upper panel of Fig.~\ref{fig:histo}), almost all the surface points are significantly polarized. In contrast, in this close to the limb scattering geometry there is only a minority of surface points with $P<1$\,\% and there is a significantly larger fraction of points of the field of view showing strong polarization signals with $P>3$\,\%. As a result, the mean fractional linear polarization amplitude at the maximum spatial resolution of the 3D model is now $\overline{P}_{\mu=0.4}\approx 1.5\,\%$.

Fig.~\ref{fig:clprofs} shows, for the case of a close to the limb LOS with $\mu=0.4$, the synthetic 
$Q/I$ and $U/I$ profiles computed at each $Y$-point without accounting for any spatial degradation; the $Y$-direction here can be assumed to be along the spectrograph's slit, assuming it to be oriented parallel to the nearest limb (i.e., the $Y$ axis). Of great interest is to compare the top panels (the zero-field reference case) with the bottom panels (where the Hanle effect of the model's magnetic field has been taken into account). In both cases the $Q/I$ signals are mainly negative, as expected from the behavior of the radiation field anisotropy ($J^2_0/J^0_0$) illustrated in Fig.~\ref{fig:jkq}. However, the polarization amplitudes are larger in the absence of magnetic fields, with a sizable factor of difference in the atmospheric regions where the magnetic field of the 3D model is relatively strong and inclined (e.g., see the region around $Y=15$\,Mm). The action of the Hanle effect can also be clearly seen in the $U/I$ signals (right panels).

The results shown above indicate that at high spatial resolution we should expect conspicuous linear polarization signals in the Ly\,$\alpha$ line core, with amplitudes often larger than $1$\,\%. These $Q/I$ and $U/I$ signals are produced by anisotropic radiation pumping in the upper solar chromosphere and they are modified, via the Hanle effect, by the vector magnetic field of the model's TR. In the following section, we turn our attention to the case of spatially averaged observations.

\section{The spatially averaged signals}  \label{sec:unresolved}

\begin{figure}[t]
\centering
\includegraphics[width=8cm]{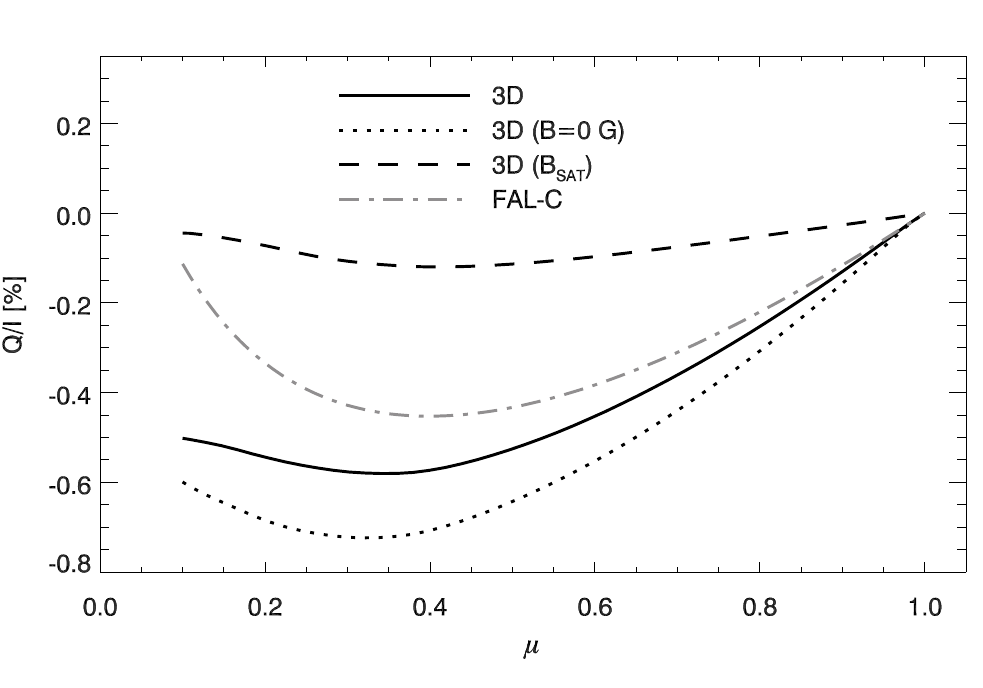}
\caption[]{
The calculated CLV of the $Q/I$ signal that results from spatially averaging the Stokes $I$ and $Q$ line-center signals at each $\mu$ location, averaged over all the LOS azimuths $\chi$, with (solid curve) and without (dotted curve) the Hanle effect produced by the model's magnetic field. The dashed curve corresponds to the Hanle effect saturation regime. The dash-dotted curve shows the calculated CLV in the semi-empirical model C of \citet{fontenla-avrett-loeser}. 
}
\label{fig:clvav}
\end{figure}

\begin{figure}[t]  
\centering
\includegraphics[width=8cm]{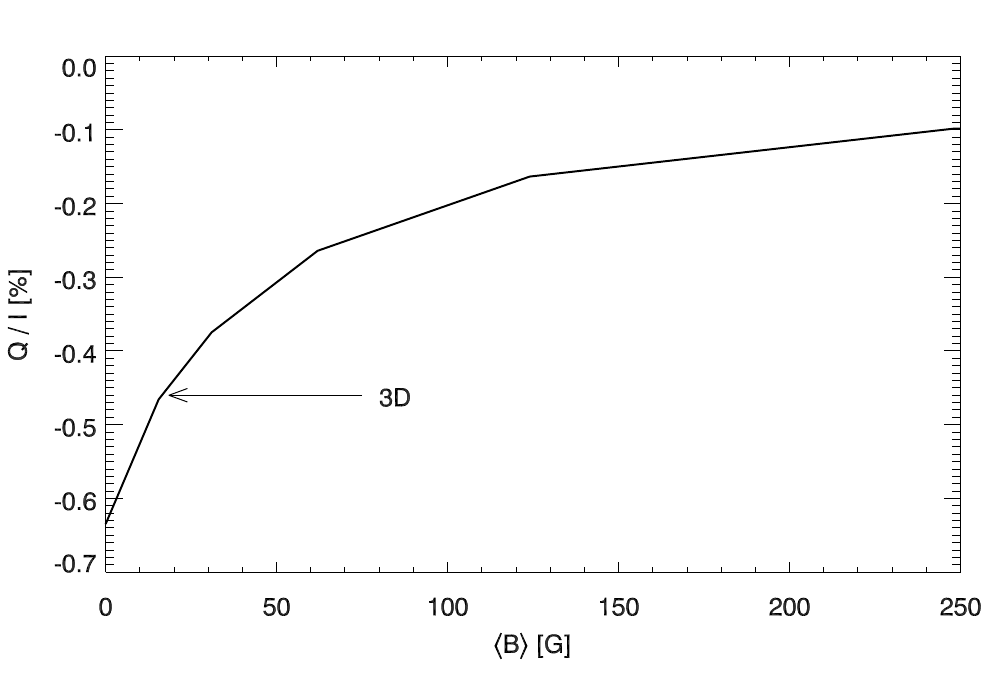}
\caption[]{
For a LOS with $\mu=0.4$ the figure shows the spatially averaged $Q/I$ amplitude vs. the mean magnetic field strength of the model's transition region. The arrow indicates the $Q/I$ signal corresponding to the original 3D model.
}
\label{fig:bsens}
\end{figure}

The solar atmosphere is a highly inhomogeneous and dynamic medium, much more complex than the idealization of a 1D, static, plane-parallel semi-empirical model atmosphere. In a 1D model atmosphere with axial symmetry around the local vertical  
the only possibility to break the axial symmetry of the incident radiation field at each point within the medium is by the presence of a magnetic field inclined with respect to the solar radius vector through the point under consideration \citep[see][and more references therein]{ishikawa14}. In the real solar atmosphere we have, in addition, the symmetry breaking produced by the horizontal inhomogeneities of the atmospheric plasma, in its density, velocity and kinetic temperature. These (non-magnetic) causes of symmetry breaking produce $Q/I$ and $U/I$ signals which could be confused with those resulting from the Hanle effect. 

In this paper we represent the complexity of the TR of the Sun by a single snapshot taken from the three-dimensional MHD simulation described in Section~\ref{sec:formulation}. One possible way to obtain quantitative information on the magnetization of the solar TR is to study the statistical properties of the Ly$\alpha$ polarization signals, and their relation with the physical quantities that describe the 3D atmospheric model. The first step towards this goal is to consider the average polarization signals and how they can be exploited to infer information on the average magnetic field strength in the TR. In this section, we study the sensitivity of the emergent $Q/I$ and $U/I$ line-center signals to the model's magnetic field, considering spatially averaged synthetic observations. As shown below, such spatially-averaged signals can be used to obtain quantitative information on the average magnetization of the solar TR.

If not stated otherwise, the results of this section 
are shown for a LOS with azimuth $\chi=0^{\circ}$, i.e., with the $Y$-axis parallel to the nearest solar limb which is also our reference direction for Stokes $Q$. In order to simplify the notation,  in what follows
the spatially averaged quantities $\langle Q\rangle/\langle I\rangle$, $\langle U\rangle/\langle I\rangle$, and $\langle P\rangle$ will be denoted by $Q/I$, $U/I$, and $P$.

\subsection{Spatial variability of the Ly\,$\alpha$ signals}  \label{ssec:spatvar}

If the spatial resolution of the observation is reduced, the emergent Stokes profiles $I$, $Q$, and $U$ are averaged over the corresponding resolution element. At the disk center, the $Q/I$, $U/I$, and $P$ signals are therefore generally degraded because of cancellation of the positive and negative signals coming from different regions of the atmosphere. Let us consider a square resolution element of size $R\times R$ located at the solar disk center. In Fig.~\ref{fig:char} we show the variation with $R$ of the total polarization signal $P$ at the Ly\,$\alpha$ line center. This result has been obtained statistically, after computing the emergent Stokes profiles for many possible realizations of the resolution element position. As seen in the figure, at $R=3''$ the total polarization line-center signal decreases by a factor two with respect to the maximum resolution value. Three seconds of arc can therefore be regarded as a characteristic spatial scale of the polarized structures seen at the center of the Ly\,$\alpha$ line. We can therefore argue that unless the spatial resolution of the observation is significantly better than 3 arcseconds, the polarization of the emergent radiation will be a mixture of signals coming from locations having significantly different plasma structures. Likewise, we can expect that for inferring the average properties of the solar TR, the spatial resolution of a single observation should be  much worse than 3 arcseconds, so that the influence of the local inhomogeneities is sufficiently reduced.

For the case of a disk-center observation, the line-center signal $P$ tends to zero in the limit of no spatial resolution. For any  LOS with $\mu<1$, the spatially averaged signals will be generally non-zero and converging, in the limit of no spatial resolution, to the values given by the center-to-limb variation (CLV). This applies to the unmagnetized case, but as shown in Fig.~\ref{fig:clvav} it happens also when we take into account the Hanle effect produced by the weak magnetic field of the chosen 3D model atmosphere. 

\subsection{Center-to-limb variation}  \label{ssec:clv}

The CLV of the emergent spectral line radiation contains useful information on the structuring of the solar atmosphere. In quiet regions of the solar surface the observed CLV of the Stokes-$I$ profile of the hydrogen Ly\,$\alpha$ line is known to be negligible \citep[e.g.,][]{roussel-dupre-1982}. However, radiative transfer calculations in the semi-empirical model C of \citet{fontenla-avrett-loeser} (hereafter, FAL-C model), show that the CLV of the $Q/I$ line-center signal is rather significant \citep[cf. the Fig.~2 of][]{jtb-stepan-casini11}. Here we study the CLV of the spatially-averaged $Q/I$ line center signals in the chosen 3D model.

The results are shown in Fig.~\ref{fig:clvav} for the case of azimuthally averaged line of sights of inclination $\mu$. Only the $Q/I$ line-center signal is shown because, as expected, the spatially averaged $U/I$ signal is practically zero for all heliocentric angles. Interestingly, the behavior of the resulting CLV of the line-center $Q/I$ signal is qualitatively similar to that found in the FAL-C model (see the dash-dotted line in Fig.~\ref{fig:clvav}). The largest fractional polarization amplitude is found around $\mu\approx 0.4$ and it decreases towards larger heliocentric angles. In contrast to the CLV behavior found in 1D models, the decrease of the $Q/I$ line-center amplitude when approaching the limb of the 3D atmospheric model 
is not so significant and in the chosen 3D model the $Q/I$ signal is still sizable at near-limb locations.

Without spatio-temporal resolution, the resulting $Q/I$ signals drop below the 1\% level, with amplitudes similar to those found in 1D semi-empirical models. Therefore, their measurement requires advanced instrumentation, capable of reaching a polarimetric sensitivity of at least 0.1\%. Via the Hanle effect, the spatially-averaged $Q/I$ signals are clearly sensitive to the model's magnetic field (see Fig.~\ref{fig:clvav}).

\subsection{Magnetic sensitivity of the average spectrum}  \label{ssec:magsens}

Following the approach introduced in Section~4 of \citet{stepan12}, we study the magnetic sensitivity of the spatially-averaged polarization amplitudes in the 3D snapshot model atmosphere. We define the quantity $\langle B\rangle$ as the average magnetic field strength at the points where the optical depth at the Ly\,$\alpha$ line center is unity for the case of a disk-center observation. Given the ``height of formation" of the Ly\,$\alpha$ line center, $\langle B\rangle$ can also be understood as the average magnetic field strength in the TR. We follow the approach of \citet{stepan12} and multiply the magnetic field strength at each grid point of the 3D model by a dimensionless factor $f\ge 0$. In this way, we can study the dependence of the line polarization on the average field strength $f\langle B\rangle$. The result for the $Q/I$ line center amplitude is shown in Fig.~\ref{fig:bsens} for a near-limb observation with $\mu=0.4$.\footnote{The variation of $U/I$ is not shown because this quantity remains practically zero for all values of $f$.} As expected, the maximum polarization signal is found in the unmagnetized case ($f=0$). The arrow in the figure indicates the actual value of the average TR magnetization (the $f=1$ case) which is $\langle B\rangle \, {=} \, 15$\,G. The $Q/I$ polarization signal gradually decreases as the magnetic field increases. In contrast to the 2D result shown in Fig.~4 of \citet{stepan12}, in the 3D model the $Q/I$ signal is lower and its maximum amplitude is more similar to that found in the FAL-C unmagnetized model.

\subsection{Detectability of the signals}  \label{ssec:tara}

\begin{figure}[t]  
\centering
\includegraphics[width=8cm]{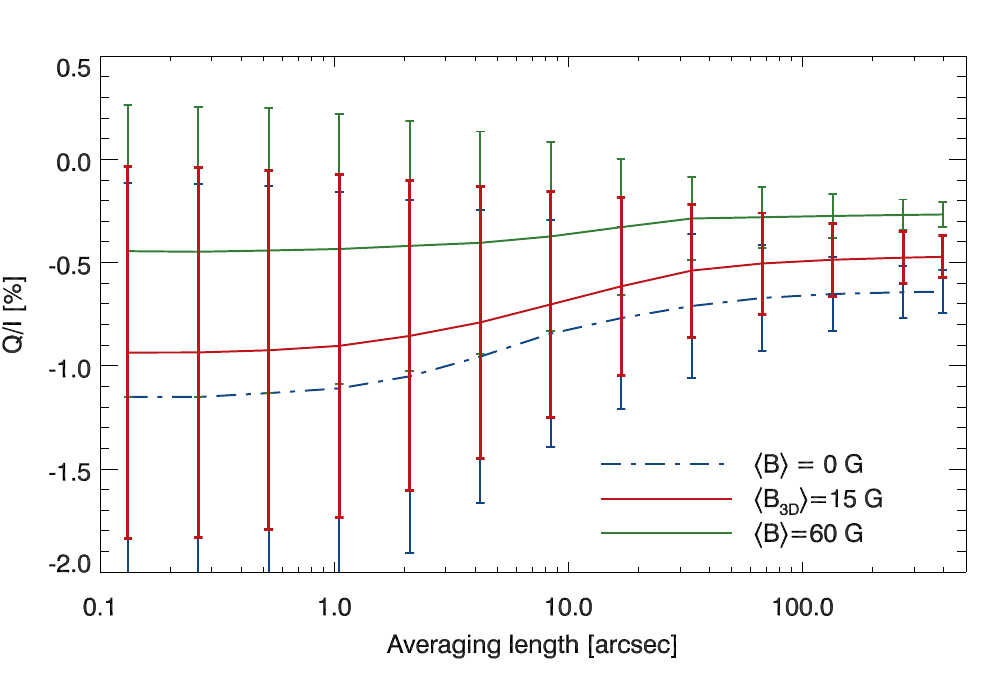}
\caption[]{
Dependence of the slit-integrated $Q/I$ amplitude 
on the averaging length along the spectrograph's slit, with the slit located at $\mu=0.4$ and oriented 
parallel to the model's solar limb. The figure shows the 
polarization signal modified by the Hanle effect of the model's magnetic field (red solid curve), the solution without the magnetic field (dash-dotted curve), and the solution obtained with the magnetic field strength increased by a factor 4 at every point of the atmospheric model (green solid curve). The error bars indicate the standard deviation of the $Q/I$ amplitudes calculated from the statistics of randomly positioned slits.
}
\label{fig:bdet}
\end{figure}

\begin{figure}[t]  
\centering
\includegraphics[width=8cm]{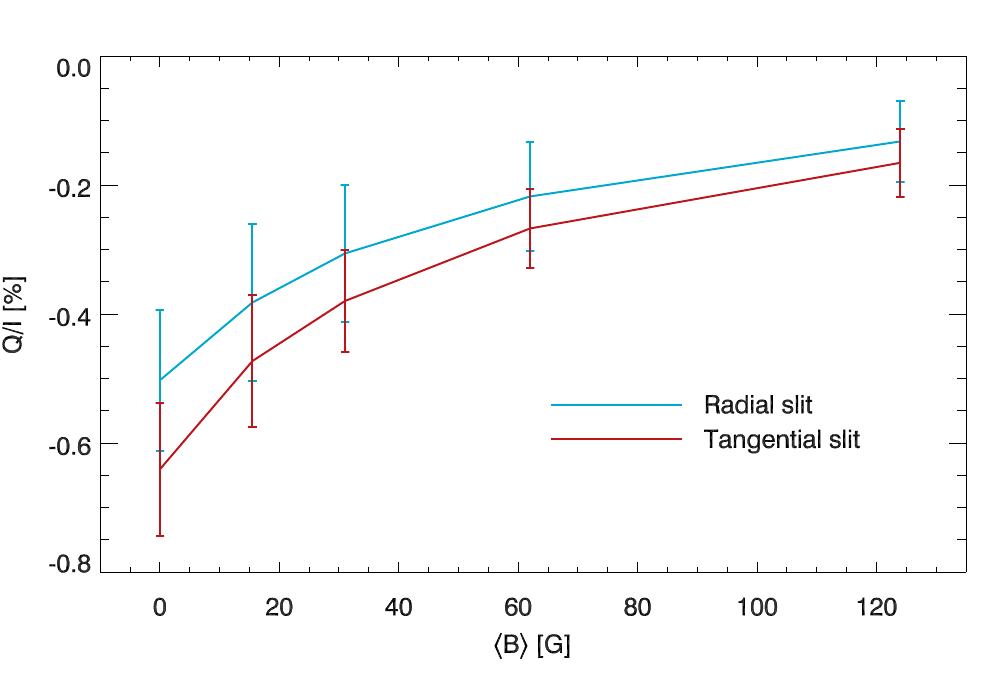}
\caption[]{
Amplitude of the $\langle Q\rangle/\langle I\rangle$ signal, integrated along the $400''$ slit, as a function of the averaged magnetic field strength $\langle B\rangle = f\langle B_{\rm 3D}\rangle$ of the model's transition region. Blue curve: Slit oriented radially and extending from $\mu=0.1$ to $\mu=0.8$. Red curve: Slit oriented tangentially with the slit center located at $\mu=0.4$.
}
\label{fig:bdettara}
\end{figure}

The Chromospheric Ly\,$\alpha$ Spectropolarimeter (CLASP) has been designed for measuring the core of the Stokes $I$, $Q$ and $U$ profiles using a spectrograph with a slit of $400''$, with a spatial resolution of about $3''$, a spectral resolution of 0.1\,\AA\ and a polarimetric sensitivity of $0.1\%$ at the $3\sigma$ level \citep[see][]{kobayashi12,kano12,kubo14}. In order to reach this level of polarimetric sensitivity it will be necessary to significantly sacrifice the temporal resolution and to partially average the observed 
Stokes $I$, $Q$, and $U$ signals along the direction of the spectrograph's slit, in order to increase the signal-to-noise ratio.
Nevertheless, in addition to the important goal of measuring for the first time scattering polarization in a FUV line (hydrogen Ly\,$\alpha$) 
whose core originates in the solar TR, CLASP can still aim at constraining the average magnetization of the TR by partial spectrum averaging along the spectrograph slit.

An important question to consider is whether the spectrograph's slit, which selects solar light along a distance of $400"$ 
on the solar disk (i.e., almost one half of the apparent solar radius), should be oriented radially or tangentially with respect to the solar limb, in order to maximize the chances of detection of the expected scattering  polarization in Ly\,$\alpha$ and the scientific output of the CLASP experiment. We dedicate this section to analyze the pros and cons of both options. Although both options seem to be suitable for detecting the scattering polarization signals, we have some reasons to prefer the radial option (see below).

We begin our analysis by considering the case of tangential orientation of the spectrograph's slit, with its center located at the heliocentric angle $\mu=0.4$, i.e., at an on-disk position where the spatially-averaged $Q/I$ signal is maximal (cf. Fig.~\ref{fig:clvav}). We have studied the dependence of the spatially-averaged $Q/I$ line-center signal on the length $L$ (in arc seconds) along the slit's direction over which we average the computed Stokes $I$, $Q$ and $U$ signals. We have done this analysis by averaging the individual signals corresponding to a large number of randomly positioned virtual slits on top of the 3D model grid.
This has been done by selecting randomly the $[X,Y]$ point that corresponds to the first point of the slit. The average is then obtained by integration of the $I$, $Q$, and $U$ signals along the randomly selected slit. If the virtual slit crosses the model's  boundary, then the integration continues along the same direction but starting at the opposite boundary, with a random position along this boundary. This way, it is possible to obtain slit-integrated signals for a large number of slits that are longer than the model's  dimensions (which is about $33''\times 33''$). We have applied this procedure for averaging lengths between $L=0.2''$ and $L=400''$. The results are shown  in Fig.~\ref{fig:bdet} for three different degrees of the model's magnetization. It is important to emphasize that the spatialy-averaged $Q/I$ signal depends on $L$: stronger signals are found, on average, for smaller lengths $L$. This is due to the fact that $Q/I$ depends on correlations between $Q$ and $I$ at particular pixels. In our 3D model, the regions with lower line intensity are, on average, more polarized than the high-intensity regions. Consequently, stronger signals are obtained for smaller lengths $L$. The spread of the resulting $Q/I$ signals is indicated by the error bars in the figure, which correspond to the standard deviation of each $Q/I$ value for any given length $L$. As expected, the larger $L$, the smaller the spread of the values. Averaging over the whole $400''$ distance of the CLASP spectrograph's  slit would provide a $2\sigma$ test for detection of the zero magnetization versus the actual model magnetization ($\langle B\rangle=15$\,G) of the 3D model. Similarly, $L=30''$ would be sufficient for a test distinguishing between a non-magnetized and a  strongly magnetized ($f\langle B\rangle=60$\,G) transition region, with a similar level of confidence.

Fig.~\ref{fig:bdettara} shows, both for the radial and tangential slit options, 
how the spatially-averaged (including averaging along the whole length of the spectrograph's slit) 
$Q/I$ line-center signal of the emergent Ly\,$\alpha$ radiation from the 3D model 
varies with the average magnetization value of the model's TR. 
For each average magnetization, $f\langle B\rangle$, a very large number of $Q/I$ signals 
were obtained by a method analogous to the one described in the previous paragraph. In the radial-slit option case, the fact that the $Q/I$ signal varies with the heliocentric angle (see Fig.~\ref{fig:clvav}) has been taken into account. The amplitude of the $Q/I$ signal predicted for the tangential configuration is slightly larger than in the radial case, but the difference is not very important and both slit orientations appear to be suitable for detecting the Ly\,$\alpha$ scattering polarization.
   
Our reasons in favor of the radial slit option for a sounding rocket experiment such as CLASP are the following. 
First, in the case of the tangential slit orientation we would have data only for a small interval of $\mu$ values, for example with 
$\mu{\approx}0.4$ at the central slit point (where the polarization amplitude is the largest). Obviously, any significant pointing error would compromise the measurement. In contrast, the radial slit option provides much more information on the CLV of the scattering 
polarization signals, and having such CLV data would facilitate reaching more reliable conclusions via detailed confrontations with radiative transfer modeling. Note that with a $400''$ radially-oriented slit one can have information from the solar limb till about $\mu=0.8$. With the radial slit option such CLV information would be available both at the line center (where the Hanle effect in Ly\,$\alpha$ operates) and in the nearby wings of the $Q/I$ profile (where effects of partial frequency redistribution and $J$-state interference come into play; see \cite{belluzzi-trujillobueno-stepan}). As shown by these authors, having in addition information on the CLV of the $Q/I$ wing signals may help to constrain (together with the observed Stokes $I$ profiles) the thermal structure of the upper chromosphere. Another reason to prefer the radial slit option is related with the fact that with the CLASP (which will measure the Ly\,$\alpha$ radiation ${\pm}0.5$ \AA\ from the line center) it will be difficult to determine with sufficient precision the zero offset of the linear polarization scale, and having the above-mentioned CLV information may help to end up with reliable conclusions in terms of the Hanle effect.

CLASP will be observing the solar atmosphere for approximately five minutes during which the structure of the dynamical TR will change to some extent. In our numerical experiments, we have only considered an instantaneous state of the solar model atmosphere corresponding to a particular snapshot of the time-dependent simulation. Furthermore, we have assumed an infinitely narrow spectrograph's slit and we have neglected any possible drift of the slit during the simulated observation. Because of the fine structuring and dynamical behavior of the solar chromosphere, CLASP cannot aim at determining the magnetic field at high spatial resolution. However, the lack of spatio-temporal resolution might actually help us to end up with observables useful to determine the mean field strength of the solar TR. Therefore, the standard deviations shown in Figs.~\ref{fig:bdet} and \ref{fig:bdettara} should be understood as being the most pessimistic estimates and the real observation will be of better accuracy.

\section{Conclusions}  \label{sec:conclusions}

The outer solar atmosphere is a highly structured and dynamic plasma, much more complex than a uniform, plane-parallel 1D configuration. Very likely, the height of the TR varies from one place in the atmosphere to another, delineating a highly crumpled surface (e.g., such as in the top left panel of Fig.~\ref{fig:jkq}). Strong spectral lines originate in this type of spatially complex plasmas, where the horizontal atmospheric inhomogeneities break the axial symmetry of the radiation field. This symmetry breaking  produces changes in the scattering polarization signals, which may compete with those caused by the Hanle effect. Therefore, the interpretation of spectropolarimetric observations in chromospheric and TR lines requires solving a complex radiative transfer problem, namely the generation of spectral line polarization in realistic 3D models of the solar atmosphere. Likewise, since the intensity profiles are practically insensitive to the magnetic field vector, the true way to validate or refute the models themselves is by confronting synthesized and measured Stokes profiles, because the magnetic field information is encoded in the spectral line polarization.     

In this paper we have extended to the full 3D case previous radiative transfer investigations of the scattering polarization in the hydrogen Ly\,$\alpha$ line, which were carried out using 1D \citep{jtb-stepan-casini11} and 2D \citep{stepan12} models of the solar atmosphere. The 3D model of the extended solar atmosphere chosen for this investigation is representative of the physical conditions of an enhanced network region, with magnetic field lines that reach chromospheric and coronal heights (see Fig.~\ref{fig:mgf}). Using the radiative transfer code PORTA developed by \citet{porta} we solved the full 3D radiative transfer problem of the generation and transfer of scattering polarization in the hydrogen Ly\,$\alpha$ line, and carried out numerical experiments to understand its magnetic sensitivity due to the Hanle effect.

The first noteworthy point is that contained in Fig.~\ref{fig:jkq}, which shows the radiation field tensors that quantify the symmetry properties of the Ly\,$\alpha$ radiation. The fractional  anisotropy of the Ly\,$\alpha$ radiation (see the top right panel) is fully negligible throughout the whole photosphere and chromosphere, but it suddenly becomes significant (of the order of 10\,\%) and predominantly negative right at the TR.\footnote{Note in the top left panel of Fig.~\ref{fig:jkq} that the model's TR practically coincides with the corrugated surface where the optical depth is unity along the LOS.}
The other panels of Fig.~\ref{fig:jkq} show that the radiation field tensors that quantify the breaking of the axial symmetry of the Ly\,$\alpha$ radiation are also negligible below the TR; i.e., in the regions where the optical depth at the line center is larger than unity.
As the optical depth at the Ly\,$\alpha$ line center becomes smaller than unity at the TR, such radiation field tensors become significant. As a result, the emergent Ly\,$\alpha$ radiation shows conspicuous scattering polarization signals, with sizable $Q/I$ and $U/I$ line-center amplitudes all over the solar disk, even in the unmagnetized reference case.

At full spatial resolution the most likely fractional polarization amplitude is 1.5\% for simulated observations at $\mu=0.4$ (close to the limb scattering geometry), while it is 0.7\% at $\mu=1$ (forward scattering geometry). The polarization signals drop below the 1\% level when the spatial resolution is degraded. For example, for a square resolution element of 3 arc seconds the most likely forward scattering polarization signal decreases by a factor two. In the limit of no spatial resolution we find that the CLV of the line-center $Q/I$ signal is qualitatively similar to that found in semi-empirical 1D models of the solar atmosphere, with the largest polarization amplitudes around $\mu=0.4$ and with $Q/I{\approx}0$ at $\mu=1$.  

Although the average magnetic field strength of the model's TR is only $\langle B\rangle \, {\approx} \, 15$\,G, the Hanle effect in the hydrogen Ly\,$\alpha$ line (whose critical Hanle field is 53 G) produces a significant depolarization (see Fig.~\ref{fig:bsens}).
For stronger magnetic fields the ensuing depolarization is larger, until it reaches the Hanle saturation limit for $\langle B\rangle \gtrsim 250$\,G. As shown in Fig.~\ref{fig:bdettara}, for the radial slit case the $\langle Q \rangle/\langle I \rangle$ line-center amplitudes are slightly smaller than in the tangential slit case, but they are equally measurable and show a similarly significant Hanle depolarization. 
In Section~\ref{ssec:tara} we give our arguments in favor of the radial slit option when observing the Ly\,$\alpha$ scattering polarization 
with the CLASP sounding rocket experiment. 

Another important conclusion is that the Hanle effect in forward scattering geometry (i.e., when observing the solar disk center) does not necessarily create linear polarization in the presence of an inclined magnetic field, as it happens in the idealized case of 1D model atmospheres. The fact that in a 3D model of the solar atmosphere the radiation field that illuminates each point of the TR plasma does not have axial symmetry with respect to the local vertical direction implies that even in the absence of a magnetic field forward-scattering processes can produce significant linear polarization signals in Ly\,$\alpha$. Interestingly, we have found and understood that in such a spatially-complex plasma the Hanle effect of an inclined magnetic field tends instead to destroy the forward-scattering polarization, as it happens when observing close to the edge of the solar disk where one reaches the case of $90^\circ$ scattering geometry. This and the previously summarized results are important for a reliable interpretation of future observations of the Ly\,$\alpha$ polarization produced by optically pumped hydrogen atoms in the chromosphere-corona transition region of the Sun.  

\acknowledgements
Financial support by the Grant Agency of the Czech Republic through grant \mbox{P209/12/P741} and project \mbox{RVO:67985815}, as well as by the Spanish Ministry of Economy and Competitiveness through projects \mbox{AYA2010--18029} (Solar Magnetism and Astrophysical Spectropolarimetry) and CONSOLIDER INGENIO CSD2009-00038 (Molecular Astrophysics: The Herschel and Alma Era) is gratefully acknowledged.
This research was also supported by the Research Council of Norway through the grant
``Solar Atmospheric Modelling'' and by the European Research Council under the European
Union's Seventh Framework Programme (FP7/2007-2013) / ERC Grant agreement Nr.
291058.
The radiative transfer simulations presented in this paper were carried out with the MareNostrum supercomputer of the 
Barcelona Supercomputing Center (National Supercomputing Center, Barcelona, Spain). We gratefully acknowledge the computing grants, technical expertise and assistance provided by the Barcelona Supercomputing Center. We are also grateful to the European Union COST action MP1104 (Polarization as a Tool to Study the Solar System and Beyond) for financing short-term scientific missions at the IAC that facilitated the development of this research project.

\end{document}